\documentclass{article}
\topmargin = - 0.5 cm
\usepackage{amsmath, amssymb, amsthm, xypic, stmaryrd,graphicx}
\newcommand{\GSC}{Guillemin--Sternberg conjecture}
\newcommand{\BCC}{Baum--Connes conjecture}

\newcommand{\car}{\circlearrowright}

\DeclareMathOperator{\coker}{coker}

\DeclareMathOperator{\ind}{index}
\DeclareMathOperator{\Preq}{Preq}
 \newcommand{\beq}{\begin{equation}}
\newcommand{\eeq}{\end{equation}} 
\newcommand{\bea}{\begin{eqnarray}}
\newcommand{\eea}{\end{eqnarray}}

\newcommand{\ca}{$C^*$-algebra} 
 \newcommand{\rep}{representation}

\newcommand{\Hs}{Hilbert space}

\newcommand{\raw}{\rightarrow}

\newcommand{\law}{\leftarrow}

 \newcommand{\Ad}{{\rm Ad}}

\newcommand{\supp}{\,{\rm supp}\,}

\newcommand{\inv}{^{-1}}

\newcommand{\er}{\eqref}
\newcommand{\gm}{\gamma} \newcommand{\Gm}{\Gamma}

\newcommand{\om}{\omega} 
 \newcommand{\B}{\mathfrak{B}}
\newcommand{\GC}{\mathfrak{C}} 
 
 \newcommand{\K}{\mathbb{K}}

 \newcommand{\g}{\mathfrak{g}}
\newcommand{\GQ}{\mathfrak{Q}}

\newcommand{\CQ}{{\mathcal Q}}

\newcommand{\C}{{\mathbb C}} \newcommand{\D}{{\mathbb D}}
\newcommand{\N}{{\mathbb N}} \newcommand{\R}{{\mathbb R}}
 \newcommand{\Z}{{\mathbb Z}}
  \makeatletter
\newskip\tempskip \def\endproof{{\parfillskip24\p@ plus\@ne
fil\@@par}\tempskip\prevdepth
\ifdim\lastskip=\z@\tempskip\z@\else\vskip-\lastskip
\ifdim\tempskip>4\p@ \tempskip.5\tempskip \else \tempskip\z@\fi\fi
\nobreak\vskip-\baselineskip\vskip-\tempskip\noindent\hbox
to\hsize{\hfill
$\blacksquare$}\par\vskip\tempskip\vskip\abovedisplayskip\@doendpe}
\makeatother \makeatletter
\newskip\tempskip \def\endiproof{{\parfillskip24\p@ plus\@ne
fil\@@par}\tempskip\prevdepth
\ifdim\lastskip=\z@\tempskip\z@\else\vskip-\lastskip
\ifdim\tempskip>4\p@ \tempskip.5\tempskip \else \tempskip\z@\fi\fi
\nobreak\vskip-\baselineskip\vskip-\tempskip\noindent\hbox
to\hsize{\hfill
$\Box$}\par\vskip\tempskip\vskip\abovedisplayskip\@doendpe}
\makeatother 

\newcommand{\Gindex}{{G\mbox{-}\mathrm{index}}}
\def\Dslash{\setbox0=\hbox{$D$}D\hskip-\wd0\hbox to\wd0{\hss\sl/\/\hss}}
\newcommand{\DS}{\Dslash}
\newcommand{\DSS}{D \!\!\!\! /}
\newcommand{\spinc}{\mathrm{Spin}^c}
\theoremstyle{plain}
\newtheorem{theorem}{Theorem}[section]
\newtheorem{lemma}[theorem]{Lemma}
\newtheorem{proposition}[theorem]{Proposition}
\newtheorem{corollary}[theorem]{Corollary}
\newtheorem{conjecture}[theorem]{Conjecture}
\theoremstyle{definition}
\newtheorem{definition}[theorem]{Definition}
\newtheorem{example}[theorem]{Example}
\newtheorem{remark}[theorem]{Remark}
\renewcommand{\D}{\DS}

\newcommand{\MGam}{M/\Gamma}
\newcommand{\GGam}{G/\Gamma}

\newcommand{\Bigwedge}{\textstyle{\bigwedge}}

\renewcommand{\B}{\mathcal{B}}
\renewcommand{\K}{\mathcal{K}}
\newcommand{\F}{\mathcal{F}}
\renewcommand{\O}{\mathcal{O}}
\newcommand{\E}{\mathcal{E}}
\begin{document}
\title{The Guillemin--Sternberg conjecture  for noncompact groups and spaces}
\author{P. Hochs \&\ N.P. Landsman \\
Radboud University Nijmegen\\
Institute for Mathematics, Astrophysics, and Particle Physics\\
Toernooiveld 1, 6525 ED NIJMEGEN,
THE NETHERLANDS\\
\texttt{hochs@math.ru.nl}, \texttt{landsman@math.ru.nl}}
\date{\today}
\maketitle

\begin{abstract}
The Guillemin--Sternberg conjecture states that ``quantisation commutes with reduction" in a specific technical setting.  So far, this conjecture has almost exclusively been stated and proved for 
{\it compact} Lie groups $G$ acting on {\it compact} symplectic manifolds, and, largely due to the use of $\spinc$ Dirac operator techniques, 
has reached a high degree of perfection under these compactness assumptions. In this paper we formulate an appropriate  Guillemin--Sternberg conjecture in the general case,
 under the main assumptions that the Lie group action 
is proper and cocompact. This formulation is motivated by our interpretation of the ``quantisation commuates with reduction" phenomenon as a special case of the
functoriality of quantisation,  and uses equivariant $K$-homology and the $K$-theory of the group \ca\ $C^*(G)$ in a crucial way. 
For example, the equivariant index - which in the compact case takes values in the \rep\ ring $R(G)$ - is replaced by the analytic assembly map - which takes values in $K_0(C^*(G))$ - familiar from the Baum--Connes conjecture in noncommutative geometry.  Under the usual freeness assumption on the action, we prove our conjecture for all Lie groups $G$ having a discrete normal subgroup $\Gm$ with compact quotient $G/\Gamma$, but we believe it is valid for all unimodular Lie groups. 
\end{abstract}
\newpage
\tableofcontents
  \section{Introduction}
In 1982 two fascinating conjectures appeared about group actions. Guillemin and Sternberg
\cite{GS82} gave a precise mathematical formulation of Dirac's idea that ``quantisation commutes with reduction" \cite{Dir}, in which they defined the former as geometric
 quantisation. As it stood, their  conjecture - which they proved under special assumptions - only made sense for actions of {\it compact} Lie groups on {\it compact} symplectic manifolds.
These compactness assumptions were left in place throughout all later refinements in the formulation of the conjecture and the ensuing proofs thereof under more general assumptions \cite{JK,Mei,meinrenkensjamaar,Met,Par1,tianzhang,Ver}.  Baum and Connes, on the other hand,
formulated a conjectural description of the $K$-theory of the reduced \ca\ $C^*_r(G)$ of a locally compact group $G$ in terms of its proper actions \cite{BC82}. Here the
emphasis was entirely on the {\it noncompact} case, as the Baum--Connes conjecture is  trivially satisfied for compact groups. The modern formulation of the conjecture in
terms of the equivariant $K$-homology of the classifying space for proper $G$-actions was given in \cite{BCH}. In this version it has now been proved for all (almost) connected groups $G$
\cite{CEN} as well as for a large class of discrete groups (see \cite{Val1}). 

Although the two conjectures in question have quite a lot in common - such as the central role played by index theory and Dirac operators,\footnote{Here the motivating role played by two closely related papers on the discrete series representations of semisimple Lie groups 
should be mentioned. It seems that Parthasarathy \cite{Parth} influenced  Bott's formulation of the \GSC\ (see below), whereas Atiyah and Schmid \cite{AS77} in part inspired the \BCC\
(see \cite{CM,Laf2}).} 
or their acute relevance  to modern physics, especially the quantisation of singular phase spaces (cf.\ \cite{LanHaag,Ober}) - the compact/noncompact divide (and perhaps also the sociological division between the communities of symplectic geometry and noncommutative geometry) seems to have
precluded much ``interaction" between them. As we shall see, however, some of the ideas surrounding
the Baum--Connes conjecture are precisely what one needs to generalize the Guillemin--Sternberg
conjecture to the locally compact case. 

Being merely desirable in mathematics, such a generalization
is actually crucial for physics. As a case in point, we mention the problem of constructing
Yang--Mills theory in dimension 4 - this is one of the Clay Mathematics Institute Millennium Prize
Problems - where the groups and spaces in question are not just noncompact, but even
infinite-dimensional! Also, the work of Dirac that initiated the modern theory of constraints and
reduction in mechanics and field theory was originally motivated by the problem  - still open - of
quantising general relativity, where again the groups and spaces are infinite-dimensional. From this perspective, the present work, in which we attempt to push the \GSC\ beyond the compactness barrier  to the locally compact situation, is merely an exiguous  step in the right direction.  

To set the stage, we display the  usual ``quantisation commutes with reduction" diagram: 
\beq 
\xymatrix{(G \car M, \omega) \ar@{|->}[r]^-{Q} \ar@{|->}[d]^{R_C}&  G\car Q(M,\omega)
  \ar@{|->}[d]^{R_Q}\\
 (M_G, \om_G)	\ar@{|->}[r]^-{Q} & Q(M_G,\om_G).} \label{qcr}
\eeq
Here $(M, \omega)$ is a symplectic manifold carrying a strongly Hamiltonian action of a Lie group $G$ (cf.\ Section \ref{sec result}), with associated 
Marsden--Weinstein quotient  $(M_G, \om_G)$, i.e.
\beq M_G=\Phi\inv(0)/G, \eeq
where $\Phi:M\raw\g^*$ is the momentum map associated to the given $G$-action (cf.\ Subsection \ref{sec group actions}).
This explains the term `$R_C$' in the diagram: \textit{C}lassical \textit{R}eduction as outlined by Dirac \cite{Dir} 
is {\it defined} as Marsden--Weinstein reduction \cite{AM,MRbook,MW,MW2}. However - and this explains both the fascination and the mystery of the field of constrained quantisation 
- the mathematical meaning of the remaining three arrows in the diagram is open to discussion: see \cite{Dir,Duv,Got,HT,landsman,Sun} for various perspectives. In any case,  $Q$ stands for \textit{Q}uantization, 
$R_Q$ denotes \textit{Q}uantum \textit{R}eduction, and all authors seem to agree that, if at all possible,  the arrows should be defined so as to make the diagram commute 
(up to isomorphism as appropriate).  We will return to the significance of this commutativity requirement below, but for the moment we just remark that it by no means 
fixes the interpretation of the arrows. 

For example, following the practice of physicists, Dirac \cite{Dir} suggested that $Q(M,\om)$ -  the quantisation of the phase space $(M, \omega)$ {\it as such} - 
should be a \Hs, which subsequently is to carry a unitary \rep\ $U$ of $G$ that ``quantises" the given canonical
 $G$-action on $M$. We assume some procedure has been selected to construct these data; see below.  
Provided that $G$ is compact, the quantum reduction operation $R_Q$ then consists in taking the $G$-invariant part $Q(M,\om)^G$ of $Q(M,\om)$.
Similarly, $Q(M_G,\om_G)$ is a \Hs\ 
without any further dressing. Now assume that $M$ and $G$ are both compact: in that case, 
the reduced space $M_G$ is compact as well, so that 
 $Q(M,\om)$ and  $Q(M_G,\om_G)$ are  typically finite-dimensional (this depends on the details of the quantisation procedure). In that case, commutativity of diagram \er{qcr} up to isomorphism just boils down to 
 the numerical equality 
\beq\dim(Q(M,\om)^G)=\dim(Q(M_G,\om_G)).\label{QR0}\eeq

This equality becomes a meaningful conjecture  once an explicit construction of the objects  $G\car Q(M,\om)$ and  $Q(M_G,\om_G)$ as specified above has been prescribed. Guillemin and Sternberg \cite{GS82}
considered the case in which
the symplectic manifold  $M$ is compact,  prequantisable, and equipped with a positive-definite complex polarization $\mathcal{J}\subset T_{\C}M$
left invariant by the given strongly Hamiltonian
$G$-action. Recall that a symplectic manifold $(M,\om)$ is called prequantisable when the cohomology class $[\om]/2\pi$ in $H^2(M,\R)$ is  integral, 
i.e.,
 lies in the image of $H^2(M,\Z)$ under the natural homomorphism $H^2(M,\Z)\raw H^2(M,\R)$.
 In that case, there exists a line bundle $L_{\om}$ over $M$ whose first Chern class $c_1(L_{\om})$ maps to $[\om]/2\pi$ under this  homomorphism; $L_{\om}$ is called the prequantisation line bundle over $M$.
 Under these circumstances,  the quantisation  operation $Q$ may be defined through geometric quantisation (cf.\ \cite{GGK} for a recent and pertinent treatment): one picks a connection $\nabla$ on $L_{\om}$ whose curvature is $\om$, and defines  the Hilbert space $Q(M,\om)$ as  the space $Q(M,\om)=H^0(M,L_{\om})$ of polarized  sections of   $L_{\om}$  (i.e.\ of sections annihilated by all $\nabla_X$, $X\in \mathcal{J}$).
This Hilbert space carries a natural unitary \rep\ of $G$ determined  by the classical data, as polarized sections of $L_{\om}$ are mapped into each other by the lift of
the $G$-action.
 Moreover, it turns out that the reduced space $M_G$ - assumed to be a manifold - inherits all relevant structures on $M$ (except, of course, the $G$-action), so that it is quantisable as well, in the same fashion. Thus (\ref{QR0}) becomes, in self-explanatory notation,  
 \beq \dim(H^0(M,L_{\om})^G) = \dim(H^0(M_G,L_{\om_G})), \label{GSC1} \eeq
  which  Guillemin and Sternberg indeed managed to prove (see also \cite{HochsHeckman}). 
   
Impressive as this is, it is hard to think of a more favourable situation for quantisation theory then the one assumed in \cite{GS82}. In  the mid-1990s,
various earlier attempts to generalize  geometric quantisation - notably in a cohomological direction - and the associated \GSC\ culminated in an unpublished proposal by  
Raoul Bott to define quantisation in terms of the (equivariant) index of a suitable Dirac operator. See, e.g., \cite{Sja}. As this definition forms the starting point of 
our generalization of the \GSC\ to the noncompact case, we consider it in some detail. (See \cite{duistermaat,friedrich, GVF,GGK} for the theory of  $\spinc$ structures and the 
associated Dirac operators). 

The first step in Bott's definition of quantisation is to canonically associate a $\spinc$ structure  $(P,\cong)$ to a given
symplectic and prequantisable manifold  $(M,\om)$ \cite{GGK,Mei}. First, one picks an almost complex structure $J$ on $M$ that is compatible with $\om$ (in that $\om(\cdot,J \cdot)$ is positive definite and symmetric, i.e.\ a metric). This $J$ canonically induces a $\spinc$ structure $P_{J}$ on $TM$ \cite{duistermaat,GGK}, but this is not the right one to use here. 
The $\spinc$ structure $P$ needed to quantise $M$ is the one obtained by twisting $P_{J}$ with the prequantisation line bundle $L_{\om}$. This means (cf.\ \cite{GGK}, App.\ D.2.7) that $P=P_{J}\times_{\ker(\pi)} U(L_{\om})$,
where $\pi: \spinc(n)\raw SO(n)$ is the usual covering projection. 
We denote the associated $\spinc$ Dirac operator by $\DS^L_M$. See for example  \cite{duistermaat,friedrich}. 
Since $M$ is even-dimensional, any Dirac  operator on $M$ worth its name decomposes in the usual way as
 \begin{equation} \label{Dmatrix}
  \DS=\left( \begin{array}{cc} 0&\DS^-\\ \DS^+& 0\end{array}
\right), 
  \end{equation}
  and we abuse notation in writing 
\beq \mathrm{index}(\DS)= \dim\ker (\DS^+)-\dim\ker (\DS^-). \label{index}
\eeq
When $M$ is compact, the Dirac operators $(\DS_M^L)^{\pm}$ determined by the $\spinc$ structure $(P,\cong)$
have finite-dimensional  kernels,  whose dimensions  define the quantisation of $(M,\om)$ as
\begin{equation}
Q(M,\om)= \mathrm{index}(\DS_M^L)\in\Z. \label{defQ}
\end{equation}
 This number turns out to be independent of the choice of the $\spinc$ structure on $M$, as long as it satisfies the above requirement, and  is
entirely determined by the cohomology class $[\om]$ (this is not true for 
the $\spinc$ structure and the  associated Dirac operator itself) \cite{GGK}.
If the symplectic manifold $(M, \omega)$ is \emph{K\"ahler} and
the  line bundle $L$ is `positive enough', then the index of $\DS_M^L$ equals the dimension of the space
$H^0(M,L)$ of holomorphic sections of the prequantum line bundle. This provides some justification for Bott's definition of quantisation.

So far,  quantisation just associates
an integer to $(M,\om)$. Bott's definition of quantisation gains in substance when a compact Lie group $G$ acts on $M$ in strongly  
Hamiltonian fashion. In that case, the pertinent $\spinc$ structure may be chosen to be $G$ invariant, and consequently the  spaces $\ker (\DS^{\pm})$  
are finite-dimensional complex $G$ modules; we denote their isomorphism classes by square brackets. 
 In this situation we write 
\begin{equation}
\Gindex(\DS)=[\ker (\DS^+)]-[\ker (\DS^-)], \label{Gindex}
\end{equation}
which defines an element of the \rep\ ring\footnote{$R(G)$ is defined as the abelian group with one generator $[L]$ for each
finite-dimensional complex \rep\ $L$ of $G$, and relations $[L]=[M]$ when
$L$ and $M$ are equivalent and $[L]+[M]=[L\oplus M]$. The tensor product
of \rep s defines a ring structure on $R(G)$.}  $R(G)$ of $G$. Thus the quantisation of $(M,\om)$ with associated $G$-action may be defined as
\begin{equation}
Q(G\car M,\om)= \Gindex(\DS_M^L)\in R(G). \label{defGQ}
\end{equation}
As before, this element only depends on $[\om]$ (and on the $G$-action, of course). 
 When $G$ is trivial, one may identify $R(e)$ with $\Z$ through the isomorphism
$$[V]-[W]\mapsto \dim(V)-\dim(W),$$ so that  (\ref{defQ})  emerges as a special case of (\ref{defGQ}).

In this setting, the Guillemin--Sternberg conjecture makes sense  
 as long as $M$ and $G$ are compact.  Namely, in diagram \er{qcr}
 the upper right corner is now construed as an element of $R(G)$, whereas the lower right corner lies in $R(e) \cong \Z$; 
 as in the original case, the geometric quantisation of the reduced space $(M_G,\om_G)$ is defined whenever that of $(M,\om)$ is.
  The quantum reduction map $R_Q:R(G)\raw\Z$ is simply defined by 
 \beq R_Q: [V]-[W]\mapsto ([V]-[W])^G := \dim(V^G)-\dim(W^G), \label{oldQR}
\eeq where
$V^G$ is the $G$-invariant part of $V$, etc.
Thus the Guillemin--Sternberg conjecture in the setting of Bott's definiton of
quantisation simply reads
\begin{equation}
 \Bigl(\Gindex\bigl(\DS^L_M\bigr)\Bigr)^G=\mathrm{index}\bigl(\DS_{M_G}^{L_G}\bigr). \label{G0}
\end{equation}
In this form, the conjecture was proved by Meinrenken  \cite{Mei}, who merely assumed that $M$ and $M_G$ are orbifolds.\footnote{Even that assumption turned out to be unnecessary \cite{meinrenkensjamaar}.} Also see \cite{GGK,JK,Par1,tianzhang,Ver} for other proofs and  further references. A step towards a noncompact version of the Guillemin-Sternberg conjecture has been taken by Paradan in \cite{Par1}, where he considers actions by compact groups on possibly noncompact manifolds. He proves that in this setting quantization commutes with reduction under certain conditions, that are met if the manifold in question is a coadjoint orbit of a semisimple Lie group, and the group acting on it is a maximal compact subgroup. Our generalisation is more or less in an orthogonal direction: we assume that the quotient of the group action is compact, rather that the group itself.

As alluded to above, using standard ideas from the context of the \BCC\ one can formulate the \GSC\ also for noncompact groups and manifolds. We specify our precise assumptions in Section \ref{sec result} below; for the moment we just mention that it seems 
impossible to even formulate the conjecture unless we assume that the strongly Hamiltonian action $G\car (M,\om)$ is {\it proper} and {\it cocompact} 
(or {\it $G$-compact}, which means that $M/G$ is compact). When this is the case, we may pass from the compact to the noncompact case by making the following replacements (or lack of these) in the formalism:
\begin{enumerate}
\item Symplectic reduction is {\it unchanged}.
\item The definition of the  $\spinc$ Dirac operator
$\DS_M^L$ associated to $(M,\om)$ is {\it unchanged}.
\item The quantisation of the reduced space (which is compact by our regularity assumptions)  is {\it unchanged}.
\item The representation ring $R(G)$ is replaced by   $K_0(C^*(G))$,\footnote{\label{forr} The use of $K_0(C^*(G))$ instead of $K_0(C_r^*(G))$ is
 actually quite unusual in the context of the \BCC; we will clarify this point in footnote \ref{crg} below.} i.e.\ the usual $K_0$ group 
of the group \ca\ of $G$.\footnote{See \cite{Blackadar,GVF,Ror,WO} for the general $K$-theory of \ca s,
see \cite{Dix,Ped} for group \ca s, and see \cite{BCH,Ros,Val1} for the $K$-theory of group \ca s.}
\item The equivariant index  $\Gindex(\DS_M^L)\in R(G)$ is replaced by 
$\mu^G_M ([\DS_M^L])\in K_0(C^*(G))$, where 
\beq
\mu^G_M: K^G_0(M)\raw K_0(C^*(G)) \label{asmap}
\eeq
 is the {\it analytic assembly map} \cite{BCH,valette, Val1}, $K^G_0(M)$ is the {\it equivariant analytical $K$-homology group} defined by $G\car M$ \cite{BCH,HigsonRoe,Kas2,Val1}, and
$[\DS_M^L]$ is the class in $K^G_0(M)$ defined by the $\spinc$ Dirac operator
$\DS_M^L$. 
\item Accordingly, the quantisation of the unreduced space $(G\car M,\om)$
is now given by 
\begin{equation}
Q(G\car M,\om)= \Gindex(\DS_M^L)\in K_0(C^*(G)), \label{defGQbis}
\end{equation}
where
\beq  \Gindex(\DS_M^L) := \mu^G_M ([\DS^L_M]) \label{jn}\eeq
purely as a matter of notation.\footnote{This notation is justified by the fact that for $G$ and $M$ compact one actually has an equality in \er{jn}, provided one identifies $K_0(C^*(G))$ with $R(G)$.}
\item The map  $R_Q: R(G)\raw\mathbb{Z}$ given by \er{oldQR}
is replaced by the  map
\beq
R_Q=\left({\textstyle \int_G }\right)_*: K_0(C^*(G))\raw\mathbb{Z} \label{newQR}
\eeq
  functorially induced by map $\int_G: C^*(G)\raw \mathbb{C}$ given by $f\mapsto \int_G  f(g)\, dg$ 
(defined on $f\in L^1(G)$ or $f\in C_c(G)$ and extended to $f\in C^*(G)$ by continuity).\footnote{\label{crg} This extension would not be defined on $C^*_r(G)$ (unless $G$ is amenable). The continuous extension to $C^*(G)$ is a trivial consequence of the fact that $\int_G$ is just the \rep\ of $C^*(G)$ corresponding to the trivial \rep\ of $G$ on $\C$ by the usual correspondence between nondegenerate \rep s of $C^*(G)$ and continuous unitary \rep s of $G$ \cite{Dix,Ped}.
} Here we make the usual identification of $K_0(\C)$ with $\Z$. 
Again, purely as a matter of notation we write this map as $x\mapsto x^G$. 
 \end{enumerate}
 With these replacements and  the notation \er{jn}, {\it our generalized Guillemin--Sternberg conjecture is formally given by its original version \er{G0}}. More precisely:
 \begin{conjecture}[Quantisation commutes with reduction] \label{con [Q,R]=0}
 Let $G$ be a unimodular Lie group, let $(M, \omega)$ be a symplectic manifold, and let $G \car M$ be a proper strongly Hamiltonian action. 
 Suppose $0$ is a regular value of the associated momentum map.
 Suppose that the action is cocompact and admits an equivariant prequantum line bundle $L$. Assume there is an almost complex structure $J$ on $M$ compatible with $\omega$. 
 Let $\DS_M^L$ be the Dirac operator on $M$ associated to $J$ and coupled to $L$, and let $\DS_{M_G}^{L_G}$ be the Dirac operator on the reduced space $M_G$, coupled to the reduced line bundle $L_G$.  Then
 \[
 \left({\textstyle \int_G}\right)_* \circ \mu_M^G\bigl[\DS_M^L \bigr] =\mathrm{index}\bigl(\DS_{M_G}^{L_G}\bigr). 
 \] 
 \end{conjecture}
 In this paper we will prove:
 \begin{theorem} \label{thm GS}
 Under the assumptions listed in Subsection \ref{sec group actions}, Conjecture \ref{con [Q,R]=0} is true.
 \end{theorem} 
A special case of the situation described in Subsection \ref{sec group actions} is the case where
$G$ is a torsion-free discrete group acting freely and cocompactly on $M$. Then Conjecture \ref{con [Q,R]=0} follows from a result of Pierrot (\cite{pierrot}, Theor\`eme 3.3.2). 
\begin{example}
Suppose $G$ is a semisimple Lie group with maximal compact subgroup $K$, and suppose $T \subset K$ is a maximal torus which is also a Cartan subgroup of $G$. Then by a theorem of Harish-Chandra, $G$ has discrete series representations. Let $\mathcal{O}_{\lambda} \subset \g^*$ 
be the coadjoint orbit of $G$ through the element $\lambda \in \mathfrak{t}^*$. Then, if $i \lambda$ is a dominant integral weight, we would expect the quantisation of $\mathcal{O}_{\lambda}$, coupled to a suitable line bundle, to be the class in $K_0(C^*(G))$ that corresponds to the discrete series representation $H_{\lambda}$ whose lowest $K$-type has highest weight $i\lambda$. In this case Conjecture \ref{con [Q,R]=0} reduces to the uninteresting equality
\[
Q\bigl( (\mathcal{O}_{\lambda})_G \bigr) = Q(\emptyset) = 0 = R_Q(Q(\mathcal{O}_{\lambda})).
\]

However, we can try to generalise Conjecture \ref{con [Q,R]=0} so that the reduction map $R_Q$ is replaced by a reduction map $R^{\mu}_Q$, which amounts to taking the multiplicity of the discrete series representation whose lowest $K$-type is the dominant weight $i\mu$ instead of the multiplicity of the trivial representation. Furthermore, we note that the symplectic reduction $(\mathcal{O}_{\lambda})^{\mu}$  of $\mathcal{O}_{\lambda}$ at the value $\mu$ is a point if $\lambda = \mu$, and empty otherwise. Therefore, we would expect that
\[
\begin{split}
Q\bigl((\mathcal{O}_{\lambda})^{\mu}\bigr) &= \delta_{\lambda \mu} \\
	&= R_Q^{\mu}(H_{\lambda}) \\
	&= R_Q^{\mu} (Q(\mathcal{O}_{\lambda})),
\end{split}
\]
where $\delta_{\lambda \mu}$ is the Kronecker delta. 
The study of Conjecture \ref{con [Q,R]=0} for 
semisimple groups is work in progress.
\end{example}
 
 The truth of our generalized \GSC\ for a special class of noncompact groups may be some justification for our specific formulation of the generalization, but in fact there is a much deeper reason why the ``quantisation commutes with reduction" issue should be stated in precisely the way we have given.  Namely,  in the above formulation {\it the \GSC\ is a special case of the 
 (conjectural) functoriality of quantisation}. This single claim summarizes a research program, of which the first steps may be found in  the papers \cite{landsman0,landsman1,landsman2}. 
 In summary, one may define a ``classical" category $\GC$ and a ``quantum" category $\GQ$, and construe the act of quantisation as a functor $\CQ:\GC\raw\GQ$. While the categories in question haven been rigorously constructed in \cite{landsman0} and \cite{Hig,Kas1}, respectively,\footnote{
 The objects of $\GC$ are  {\it integrable}  Poisson manifolds and its arrows are {\it regular}  Weinstein dual pairs; see  \cite{landsman0} for the meaning of the
 qualifiers. The arrows are composed by a generalization of the symplectic reduction procedure, and isomorphism of objects in $\GC$ turns out to be the same as Morita
 equivalence of Poisson manifolds in the sense of Xu \cite{Xu}.  The category $\GQ$ is nothing but the Kasparov--Higson category $K\! K$, whose objects are separable \ca s and
 whose sets of arrows are Kasparov's $K\! K$-groups, composed with Kasparov's intersection product.
 See  \cite{Blackadar,Hig,Kas1}.}  the existence of the functor $\CQ$ is so far hypothetical. However, the picture that emerges from  the cases where $\CQ$  has been constructed should hold in complete generality: {\it  deformation quantisation} (in the \ca ic sense first proposed by Rieffel \cite{Rie1,Rie2}) {\it is the object side of $\CQ$, whereas geometric quantisation} (in the sense of Bott as explained above) {\it is the arrow side of $\CQ$}. Moreover, in the setting of strongly Hamiltonian group actions as considered above,  the ``quantisation commutes with reduction" principle is nothing but the  functoriality of quantisation
 (in cases where the functor has  indeed been defined).\footnote{
 Let $G\car (M,\om)$ define the Weinstein dual pair $pt\law M\raw \mathfrak{g}^*$ in the usual way \cite{Wei83}, the arrow $\raw$ being given by the momentum map.  Functoriality of quantisation means that 
$\CQ(pt\law M\raw \mathfrak{g}^*)\times_{K\! K} \CQ(\mathfrak{g}^*\hookleftarrow 0\raw pt)  =  \CQ((pt\law M\raw\mathfrak{g}^*)\times_C (\mathfrak{g}^*\hookleftarrow 0\raw pt))$. 
This equality is exactly the same as \er{G0}. See \cite{landsman2}.}

\subsection*{Outline of the proof}

 We believe our generalized Guillemin--Sternberg  conjecture to be true for all unimodular Lie groups $G$, but for reasons of human frailty we are only able to prove it in this paper when $G$ has a discrete normal subgroup $\Gamma$, such that the quotient group $K:=G/\Gm$ is compact.\footnote{Such groups are automatically unimodular. 
 The \GSC\ may not hold in the non-unimodular case; see \cite{DuTu}.}
  This incorporates a number of interesting examples. Our proof is based on:
  \begin{enumerate}
\item 

 The validity of the \GSC\ in the compact case  \cite{JK,Mei,Met,meinrenkensjamaar,Par1,tianzhang,Ver};
 \item Naturality of the assembly map for discrete groups \cite{valette};
 \item Symplectic reduction in stages \cite{landsman,MMOPS,MW2};
 \item Quantum reduction in stages.
 \end{enumerate}
In this paper we show, among other things, that the \GSC\ for discrete (and possibly noncompact) groups $G$ is a consequence of the second point. For $G$ as specified in the previous paragraph, 
 naturality of the assembly map implies a $K$-equivariant version thereof.  The third and fourth items are used in an almost trivial way, namely in setting up the following 
 diagram, which provides an outline of our proof.
\begin{equation} \label{eq diag pf}
\xymatrix{
\Preq(G \car M, \omega) \ar[r]^-{\left[\DSS_M^{\,\bullet}\right]} \ar[d]^{R_C^{(\Gamma)}} & K_0^G(M) \ar[r]^-{\mu^G_M} & K_0(C^*(G)) \ar[d]^{R_Q^{(\Gamma)}} \\
\Preq(K \car M_{\Gamma}, \omega_{\Gamma}) \ar[r]^-{\bigl[\DSS_{M_{\Gamma}}^{\,\bullet}\bigr]} \ar[d]^{R_C^{(K)}} & K_0^K(M_{\Gamma}) \ar[r]^-{\mu_{M_{\Gamma}}^K} & 
															K_0(C^*(K)) \ar[d]^{R_Q^{(K)}} \\
\Preq((M_{\Gamma})_K), (\omega_{\Gamma})_K) \ar[r]^-{\bigl[\DSS_{M_G}^{\,\bullet}\bigr]} &  K_0\bigl((M_{\Gamma})_K\bigr) \ar[r]^-{\ind} & \Z
}
\end{equation}
Here the following notation is used. We write
\begin{eqnarray}
K&=&G/\Gamma; \\ M_{\Gamma}&=&M/\Gamma, 
\end{eqnarray}
as $\Gm$ is discrete (so that its associated momentum map $\Phi_{\Gm}$ is identically zero, 
whence $\Phi_{\Gm}\inv(0)=M$). 
Furthermore, $\Preq(G \car M, \omega) $ is defined relative to a {\it given} Hamiltonian action of $G$ on a symplectic manifold $(M, \omega)$, and consists
of all possible prequantisations $(L, \nabla, H)$ of this action. A necessary condition for $\Preq(G \car M, \omega)$ to be nonempty is that the cohomology class
$[\omega]/{2\pi} \in H^2(M, \R)$ be integral. (If the group $G$ is compact, this condition is also sufficient.) We make this assumption. 
Similarly, $\Preq(K \car M_{\Gamma}, \omega_{\Gamma})$
is defined {\it given} the $K$-action on $M_{\Gamma}$ induced by the $G$-action on $M$, 
and $\Preq((M_{\Gamma})_K, (\omega_{\Gamma})_K)$ is just the set of prequantisations of the symplectic manifold
\beq
\bigl((M_{\Gamma})_K, (\omega_{\Gamma})_K\bigr)\cong (M_G, \omega_G); \label{istages}\eeq
this isomorphism is a special and almost trivial case of the theorem on symplectic reduction in stages \cite{landsman,MMOPS}. The maps $R_C^{(\Gamma)}$ and $R_C^{(K)}$ 
denote Marsden--Weinstein reduction (at zero) with respect to the groups $\Gamma$ and $K$, respectively.
We define the quantum counterparts of these maps by 
\begin{eqnarray}R_Q^{(\Gamma)} &:=& \left( {\textstyle \sum_{\Gamma} }\right)_* ; \label{eq Gammared}\\
 R_Q^{(K)} &:=& \left( {\textstyle \int_{K} }\right)_* .\end{eqnarray}
Here 
$\left( \sum_{\Gamma}\right)_*:K_0(C^*(G))\raw K_0(C^*(K))$ is the map 
 functorially induced by the map  $\sum_{\Gamma}: C^*(G)\raw C^*(G/\Gm)$ given by
 \beq \bigl({\textstyle \sum_{\Gamma}}f \bigr) (\Gamma g)=\sum_{\gm\in\Gm}f(\gm g), \eeq
initially defined on $f\in C_c(G)$; see \cite{Green77} for the continuity of this map.\footnote{This map can more generally be defined for any closed 
normal subgroup $N$ of $G$, cf.\ Appendix \ref{app nat}.}  
Finally,  the maps involving the symbol $[\D^{\bullet}]$ are defined by taking the $K$-homology class of the Dirac operator coupled to a given prequantum line
bundle,
(as outlined above and as explained in detail in the main body of the paper below). 
 Thus the commutativity of the upper part of diagram \er{eq diag pf} is the equality 
\beq
\mu_{M_{\Gamma}}^K\bigl[\DS_{M_{\Gamma}}^{L_{\Gamma}}\bigr] = R_{Q}^{(\Gamma)} \bigl( \mu_M^G [\DS_M^L]\bigr),
\eeq
whereas commutativity of the lower part yields
\beq
\ind \DS_{(M_{\Gamma})_K}^{(L_{\Gamma})_K} = R_Q^{(K)} \bigl( \mu_{M_{\Gamma}}^K \bigl[ \DS_{M_{\Gamma}}^{L_{\Gamma}} \bigr]\bigr).
\eeq

It is easily shown that
\beq
{\textstyle \int_K} \circ {\textstyle \sum_{\Gm}} ={\textstyle \int_G}, \label{easy} \eeq
so that by functoriality of $K_0$ one has
\beq
R_Q^{(K)}\circ R_Q^{(\Gm)}=R_Q^{(G)}, \label{qrs}
\eeq
with $R_Q^{(G)}=R_Q$ as in \eqref{newQR}.
Using \er{qrs} and 
\beq R_C^{(K)}\circ R_C^{(\Gm)}=R_C^{(G)}:=R_C, \label{crs} \eeq
which is a mere rewriting of \er{istages}, 
we see that the outer diagram in \er{eq diag pf} is equal to 

\begin{equation} \label{outer}
\xymatrix{
\Preq(G \car M, \omega) \ar[d]^{R_C} \ar[r]^-{Q} & K_0(C^*(G))  \ar[d]^{R_Q} \\
\Preq(M_G, \omega_G) \ar[r]^-{Q} & \Z .}
\end{equation}
Clearly,  commutativity of \er{outer} is precisely the commutativity of diagram \er{qcr}
with the post-modern meaning we have given to its ingredients. 
Since diagram \er{outer} commutes when the two inner diagrams in diagram 
\eqref{eq diag pf} commute,  the latter would prove our  generalized \GSC. 
Now the lower diagram commutes by 
 the validity of the \GSC\ for compact $K$, whereas the upper diagram
 decomposes as 
  \beq \label{eq upper diag}
  \xymatrix{
 \Preq(G \car M, \omega) \ar[r]^-{\left[\DSS_M^{\, \bullet}\right]} \ar[d]^{R_C^{(\Gamma)}} & K_0^G(M) \ar[r]^{\mu^G_M}  \ar[d]^{V_{\Gamma}}& 
 			K_0(C^*(G)) \ar[d]^{R_{Q}^{(\Gamma)} } \\
\Preq(K \car M_{\Gamma}, \omega_{\Gamma}) \ar[r]^-{\bigl[\DSS_{M_{\Gamma}}^{\, \bullet}\bigr]}   & K_0^K(M_{\Gamma}) \ar[r]^{\mu_{M_{\Gamma}}^K} & K_0(C^*(K)), 
  } 
  \eeq
  where $V_{\Gamma}$ is a map  defined in Subsection \ref{subsec L^2} (the $V$ stands for \textit{V}alette, who was the first to write this map down in a more special context). 
  Verifying the commutativity of the two inner diagrams of diagram \eqref{eq upper diag}, then, forms the main burden of our proof.
   
The commutativity of the right-hand inner diagram
 follows from a generalization of the naturality of the  assembly map for discrete groups as proved by Valette \cite{valette} to possibly nondiscrete groups.   This is dealt with in Appendix \ref{app nat}. 
 The commutativity of the left-hand inner diagram is Theorem \ref{thm reduction}:
 \beq
 V_{\Gamma}[\DS_M^L] = \bigl[\DS_{M_{\Gamma}}^{L_{\Gamma}}\bigr] .\label{t28}
 \eeq
  The proof of this result occupies Sections \ref{sec diff ops} and \ref{sec Dirac operators}.
 
In Section \ref{sec diff ops}, we compute the image under the map $V_{\Gamma}$ of a $K$-homology class associated to a general equivariant, elliptic, symmetric, first order differential operator $D$ on a
$\Gamma$-vector bundle $E$ over a $\Gamma$-manifold $M$. If the action of $\Gamma$ on $M$ is free, as we assume, then the quotient space $E/\Gamma$ defines a vector bundle over
$M/\Gamma$. The operator $D$ induces an operator $D^{\Gamma}$ on this quotient bundle. It turns out that the homomorphism $V_{\Gamma}$ maps the class associated to $D$ to
the class associated to $D^{\Gamma}$.

In Section \ref{sec Dirac operators}, we show that if $\DS_M^L$ is the Dirac operator on a symplectic manifold $M$, coupled to a prequantum line bundle $L$, then the operator
$\bigl(\DS_M^L \bigr)^{\Gamma}$ from the previous paragraph is precisely the Dirac operator on the quotient $M/\Gamma$ coupled to the line bundle $L/\Gamma$.  
 
 As an {\it encore}, in Section \ref{Abeliancase} we give an independent proof of our generalized \GSC\ for the case that $G$ is discrete and abelian. 
 This proof, based on a paper by Lusztig \cite{Lusztig} (see also \cite{BCH},  pp.\ 242--243) gives considerable insight in the situation. 
 It is based on an explicit computation of the image under $\mu^{\Gamma}_M$ of a
$K$-homology class $[D]$ associated to a $\Gamma$-equivariant elliptic differential operator $D$ on a $\Gamma$-vector bundle $E$ over a $\Gamma$-manifold $M$. 
Because in this
case $C^*(\Gamma) \cong C(\hat \Gamma)$ (with $\hat \Gamma$ the unitary dual of $\Gamma$), this image corresponds to the formal difference of two
equivalence classes of vector bundles over $\hat \Gamma$. These bundles are described as the kernel and cokernel of a `field of operators' 
$\bigl(D_{\alpha}\bigr)_{\alpha \in \hat \Gamma}$ on a `field of vector bundles' $\bigl( E_{\alpha} \to M/\Gamma\bigr)_{\alpha \in \hat \Gamma}$. The operators $D_{\alpha}$
and the bundles $E_{\alpha}$ are constructed explicitly from $D$ and $E$, respectively. The quantum reduction of the class $\mu^{\Gamma}_M[D]$ is the index of the operator
$D_1$ on $E_1 \to M/\Gamma$, where $1 \in \hat \Gamma$ is the trivial representation. Because $D_1$ is the operator $D^{\Gamma}$ mentioned above, the \GSC\ 
follows from the computation in Section \ref{sec Dirac operators}.

Finally, in Section \ref{sec example} we check the discrete  abelian case in an instructive explicit computation. 
We will see that the quantisation of the action of $\Z^2$ on $\R^2$ corresponds to a certain line bundle over the two-torus 
$\mathbb{T}^2=\hat \Z^2$. The quantum reduction of this $K$-theory class is its rank, the integer $1$. This is also the quantisation of the reduced space 
$\mathbb{T}^2 = \R^2/\Z^2$, as can be seen either directly or by applying Atiyah-Singer for Dirac operators. Although this is the simplest example of Guillemin-Sternberg for noncompact groups and spaces, the details are nontrivial and, in our opinion, well worth spelling out.

\subsection*{Acknowledgements}
This work is supported by {\sc n.w.o.} through grant no.\ 616.062.384 for the second author's `Pionier project' {\it Quantization, noncommutative geometry, and symmetry}.

The authors would like to thank Erik van den Ban, Rogier Bos, Siegfried Echterhoff, Gert Heckman, Herv\'e Oyono-Oyono, John Roe, 
Elmar Schrohe, Alain Valette and Jan Wiegerinck for useful suggestions at various stages of this work. The authors are also indebted to the referees for several useful remarks.

\section{Assumptions and result} \label{sec result}

We now state the assumptions under which we will prove our generalised Guillemin--Sternberg conjecture, i.e.\ 
Theorem \ref{thm GS}. These assumptions are mainly used in the proof of our key intermediate result, Theorem \ref{thm reduction}, which is 
proved in Sections \ref{sec diff ops} and \ref{sec Dirac operators}.

We first fix some notation and assumptions. If $M$ is a manifold, then the spaces of vector fields and differential forms 
on $M$ are denoted by $\mathfrak{X}(M)$ and $\Omega^*(M)$, respectively. 
The symbol $\lrcorner$ denotes contraction of differential forms by vector fields.
If the manifold $M$ is equipped with an almost complex structure, then 
we have the space $\Omega^{0,*}(M)$ of differential forms on $M$ of type  $(0,*)$. 
Unless stated otherwise, all manifolds, maps and actions are supposed to be 
$C^{\infty}$.

If a vector bundle $E \to M$ is given (which is supposed to be complex unless stated
otherwise), then the space of smooth sections of $E$ is denoted
by $C^{\infty}(M, E)$. If $M$ is equipped with a measure, and the bundle $E$ carries a metric, then $L^2(M, E)$ the space of square integrable sections of $E$.
The space of differential forms on $M$ with coefficients in $E$ is denoted by $\Omega^*(M; E)$, and similarly we have the space $\Omega^{0,*}(M; E)$
for almost complex manifolds. If $F \to M$ is another vector bundle, and $\varphi: E \to F$ is a homomorphism of vector bundles, then composition with $\varphi$ gives a
homomorphism of $C^{\infty}(M)$-modules
\[
\tilde \varphi: C^{\infty}(M, E) \to C^{\infty}(M, F).
\]

If a group $G$ acts on $M$, and if $E$ is a $G$-vector bundle over $M$, then we have the canonical representation of $G$ on $C^{\infty}(M, E)$ given by 
$(g\cdot s)(m) = g\cdot s(g^{-1}m)$, for $g \in G$ and $s \in C^{\infty}(M,E)$. 
A superscript `$G$' denotes the subspace of $G$-invariant elements. Thus we obtain for example the vector spaces $C^{\infty}(M, E)^G$, 
$\Omega^{0,*}(M)^G$, etc.
The Lie algebra of a Lie group is denoted by a lower case Gothic letter, so that for example the group $G$ has the Lie algebra $\g$.

In the context of Hilbert spaces and Hilbert $C^*$-modules, we denote the spaces of bounded, compact and finite-rank operators by $\B$, $\K$ and $\F$, respectively.

 The spaces of continuous functions, bounded continuous functions, continuous functions vanishing at infinity and compactly supported continuous functions  on a topological space $X$ are denoted by $C(X)$, $C_b(X)$, $C_0(X)$ and $C_c(X)$, respectively.

\subsection{Assumptions} \label{sec group actions}

Let $(M, \omega)$ be a symplectic manifold, and let $G$ be a Lie group. 
Suppose that $G$ has a \textit{discrete}, \textit{normal} subgroup $\Gamma$, such that the quotient group $K:=\GGam$ is 
\textit{compact}.  For example, 
\begin{itemize}
\item $G=K$, $\Gamma = \{e\}$ with $K$ a compact Lie group, 
\item $G=\Gamma$ discrete, 
\item $G=\R^n$, $\Gamma = \Z^n$ so that $G/\Gamma$ is the torus $\mathbb{T}^n$, 
\end{itemize}
or direct products of these three examples. In fact, if $G$ is connected, then the subgroup $\Gamma$ must be central, 
and $G$ is the product of a compact group and a vector space. 

The assumption that $G/\Gamma$ is compact is not needed in the proof of Theorem \ref{thm reduction};
 it is only made so that we can apply the
Guillemin--Sternberg theorem for compact groups in diagram \eqref{eq diag pf}.

Suppose that $G$ acts on $M$, and that the following assumptions hold.
\begin{enumerate}

\item The action is proper.

\item The action preserves the symplectic form $\omega$.

\item The quotient space $M/G$ is compact.

\item The action is Hamiltonian,\footnote{Sometimes an action is
called `Hamiltonian' as opposed to `strongly Hamiltonian' if it admits a momentum map that is not necessarily equivariant or Poisson. We will not use this terminology;
for us the word `Hamiltonian' always means `strongly Hamiltonian'.} in the sense that there exists a map
\[
\Phi: M \to \g^*,
\]
that is equivariant with respect to the co-adjoint representation of $G$ in $\g^*$, such that for all $X \in \g$,
\[
d\Phi_X = -X_M \lrcorner \omega.
\]
Here $\Phi_X$ is the function on $M$ obtained by pairing $\Phi$ with $X$ and $X_M$ is the vector field on $M$ induced by $X$. 

\item The discrete subgroup $\Gamma$ acts freely on $M$, and the whole group $G$ acts freely on the level set $\Phi^{-1}(0)$.

\item The symplectic manifold $(M, \omega)$ admits a $G$-equivariant prequantisation. That is, there is a $G$-equivariant complex line bundle
\[
L \to M,
\]
equipped with a $G$-invariant Hermitian metric $H$, and a  Hermitian connection $\nabla$ with curvature two-form  $\nabla^2 = 2\pi i \, \omega$, which is equivariant as an operator from $C^{\infty}(M, L)$ to $\Omega^1(M; L)$.

\item There is a $G$-invariant almost complex structure $J$ on $TM$, such that
\[
B(\cdot, \cdot) := \omega(\cdot, J \cdot)
\]
defines a Riemannian metric on $M$. 

\item The manifold $M$ is  complete with respect to the Riemannian metric $B$.

\end{enumerate}

\textit{Ad 3.} Because the quotient $G/\Gamma$ is compact, compactness of $M/G$ is equivalent to compactness of $M/\Gamma$. The latter assumption is used in the proof of
Lemma \ref{lem chi surjective} (where $\Gamma$ is replaced by a general closed normal subgroup $N$), but it is not essential. Furthermore, in the definitions of $K$-homology in \cite{BCH, valette} it is assumed that the orbit spaces of the group actions involved are
compact.\footnote{$K$-homology can be defined more generally, but the compactness assumption makes things a little easier.} Finally, compactness of $M/\Gamma$ allows us to apply the Guillemin-Sternberg conjecture for compact groups and spaces to this quotient.

\textit{Ad 4.} The map $\Phi$ is called a \textit{momentum map} of the action. For connected groups, the assumption that $\Phi$ is equivariant is equivalent to the assumpion that it is a Poisson map with respect to the negative Lie-Poisson structure on $\g^*$ (see for example  \cite{landsman}, Corollary III.1.2.5 or \cite{MRbook}, \S 11.6).
Note that if $G=\Gamma$ is discrete, then the action is automatically  Hamiltonian. Indeed, $\mathrm{Lie}(\Gamma)=\{0\}$, so that the zero map is a momentum map.

\textit{Ad 5.} Freeness of the action of $\Gamma$ on $M$ implies that the quotient $M/\Gamma$ is smooth, and that a $\Gamma$-vector bundle $E \to M$ induces a vector bundle
$E/\Gamma \to M/\Gamma$. And if $G$ acts freely on $\Phi^{-1}(0)$, then it follows from de definition of momentum maps that $0$ is a regular value of $\Phi$ (Smale's lemma
\cite{Smale}), so that the reduced space $M_G$ is a smooth manifold.
If freeness is replaced by local freeness in Assumption 5, then $M/\Gamma$ and $M_G$ are \textit{orbifolds}. In that case, a quotient vector bundle $E/\Gamma \to
M/\Gamma$ can be replaced by `the vector bundle over $M/\Gamma$ whose space of smooth sections is $C^{\infty}(M, E)^{\Gamma}$ 
(see Proposition \ref{prop quotient vector bundle}).\footnote{ 
It is not a good idea to make the stronger assumption that $G$ acts freely on $M$. For in that case, $G$ must be discrete (see Remark \ref{rem discrete}).}

\textit{Ad 6. } Example \ref{ex R2} below shows that it is not always obvious if this assumption is satisfied. Equivariance of the connection $\nabla$ 
implies that the Dolbeault--Dirac operator on $M$, coupled to $L$ via $\nabla$ (Definition \ref{rem Dolbeault--Dirac op}) is equivariant.

The Kostant formula
\[
X \mapsto -\nabla_{X_M} + 2\pi i \, \Phi_X
\]
defines a representation of $\g$ in the space of smooth sections of $L$. In the literature on the Guillemin-Sternberg conjecture, it is usually assumed that the action of $G$ on $L$ is such that the corresponding 
representation of $\g$ in $C^{\infty}(M, L)$ is given by the Kostant formula. Then, if the group $G$ is connected, the connection $\nabla$ 
satisfies $g\nabla_v g^{-1} = \nabla_{g\cdot v}$ for all $g \in G$ and $v \in \mathfrak{X}(M)$. This property is equivalent to equivariance of the connection $\nabla$ in the sense of assumption 6.

If the manifold $M$ is simply connected and the group $G$ is discrete, then Hawkins \cite{hawkins} gives a procedure to lift the action of $G$ on $M$ to a projective action on the trivial line bundle over $M$, such that a given connection is equivariant. Under a certain condition (integrality of a group cocycle), this projective action is an actual action.

\textit{Ad 7.} An equivalent assumption is that there exists a $G$-invariant Riemannian metric on $M$. (See e.g.\ \cite{GGK}, pp. 111-112.)

\textit{Ad 8.} This assumption implies that the Dirac operator on $M$ is essentially self-adjoint on its natural domain (see Subsection \ref{sec Dirac oper}).  

\medskip

\noindent The assumptions and notation above will be used in this section and in Section \ref{sec Dirac operators}.
In Section \ref{sec diff ops}, we will work under more general assumptions. 

\begin{remark}
If the group $G$ is compact, then some of these assumptions are always satisfied. First of all, the action is automatically proper. Furthermore, if the
cohomology class $[\omega] \in H^2_{\mathrm{dR}}(M)$ is integral, then a prequantum line bundle exists, and the connection can be made equivariant by averaging over $G$. 
Also, averaging over $G$ makes any Riemannian metric $G$-invariant, so that assumption 7 is also satisfied. And finally, since $M/G$ is compact, so is $M$. In particular, 
$M$ is complete, so assumption 8 is satisfied.
\end{remark}

\begin{remark} \label{rem discrete}
If the action of $G$ on $M$ is (locally) free and Hamiltonian, and $M/G$ is compact, then $G$ must be \textit{discrete}.
Indeed, if the action is locally free then by Smale's lemma the momentum map $\Phi$ is a submersion, and in particular an open mapping.
And since it is $G$-equivariant, it induces
\[
\Phi^G: M/G \to \g^*/\Ad^*(G),
\]
which is also open. So, since $M/G$ is compact, the image
\[
\Phi^G(M/G) \subset \g^*/\Ad^*(G)
\]
is a compact open subset. Because $\g^*/\Ad^*(G)$ is connected, it must therefore be compact. This, however, can only be the case (under our assumptions) when $G$ is discrete.\footnote{If $G=K$ is a compact connected Lie group, then $\mathfrak{k}^*/\Ad^*(K)$ is a Weyl chamber.}
 Indeed, we have
\[
\Ad^*(G) \cong \Ad^*(K) \subset GL(\mathfrak{k}^*) \cong GL(\g^*).
\]
So $\Ad^*(G)$ is compact, and $\g^*/\Ad^*(G)$ cannot be compact, unless $\g^*=0$, i.e.\ $G$ is discrete.
\end{remark}

\begin{example}\label{ex R2}
Let $M = \C$, with coordinate $z = q+ip$, and the standard complex structure. We equip $M$ with the symplectic form $\omega = dp \wedge dq$.

Consider the group $G = \Gamma = \Z +i\Z \subset \C$. We let it act on $M$ by addition:
\[
(k+il)\cdot z = z+k+il,
\]
for $k, l \in \Z$, $z \in \C$.

Consider the trivial line bundle $L = M \times \C \to M$. We define an action of $\Z + i\Z$ on $L$ by letting the elements $1, i \in \Z + i\Z$ act as follows:
\[
\begin{split}
1\cdot(z, w) &= (z+1, w); \\
i\cdot(z, w) &= (z+i, e^{-2\pi i z}w), 
\end{split}
\]
for $z, w \in \C$. Define a Hermitian metric $H$ on $L$ by
\[
H\left((q+ip, w), (q+ip, w')\right) = e^{2 \pi (p-p^2)}w\bar w'. 
\]
This metric is $\Z + i\Z$-invariant, and the connection
\[
\nabla = d + 2\pi i \,  p \,  dz + \pi \, dp
\]
is Hermitian, $\Z+i\Z$-invariant, and has curvature form $2\pi i \, \omega$.

The details of this example are worked out in Section \ref{sec example}, where we also give some motivation for these formulae. 
\end{example}

\begin{example}
Suppose $(M_1, \omega_1)$ is a compact symplectic manifold,  $K$ is a compact Lie group, and let a proper Hamiltonian action of $K$ on $M_1$ be given. Let $\Phi$ be the
momentum map, and suppose $K$ acts freely on $\Phi^{-1}(0)$. Suppose $[\omega]$ is an integral cohomology class. 
(These assumptions are made for example by Tian \&\ Zhang \cite{tianzhang}.)
Let $\Gamma$ be a discrete group acting properly and freely on a symplectic manifold $(M_2, \omega_2)$, leaving $\omega_2$ invariant. Suppose that $M_2/\Gamma$ is compact,
and that there is an equivariant prequantum line bundle over $M_2$.
Then the direct product action of $K \times \Gamma$ on $M_1 \times M_2$ satisfies the assumptions of this section.
\end{example}

\subsection{The Dirac operator} \label{sec Dirac oper}

The prequantum connection $\nabla$ on $L$ induces a differential operator
\[
\nabla: \Omega^k(M; L) \to \Omega^{k+1}(M; L),
\]
such that for all $\alpha \in \Omega^{k}(M)$ and $s \in C^{\infty}(M, L)$,
\[
\nabla(\alpha \otimes s) = d\alpha \otimes s + (-1)^k \alpha \wedge \nabla s.
\]
Let $\pi^{0, *}$ be the projection
\[
\pi^{0, *}: \Omega^{*}_{\C}(M; L) \to \Omega^{0, *}(M; L).
\]
Composing the restriction to $\Omega^{0, q}(M; L)$ of the complexification of $\nabla$ with $\pi^{0, *}$, we obtain a differential operator
\[
\bar \partial_L := \pi^{0, *} \circ \nabla: \Omega^{0, q}(M; L) \to \Omega^{0, q+1}(M; L).
\]
Let 
\[
\bar \partial_L^*:  \Omega^{0, q+1}(M; L) \to \Omega^{0, q}(M; L)
\]
be the formal adjoint of $\bar \partial_L$ with respect to the $L^2$-inner product on compactly supported forms. 
\begin{definition}[Dolbeault--Dirac operator] \label{rem Dolbeault--Dirac op}
The \textit{Dolbeault--Dirac operator} on $M$, coupled to $L$ via $\nabla$ is the differential operator
\[
\DS_M^L := \bar \partial_L + \bar \partial_L^*: \Omega^{0, *}(M; L) \to \Omega^{0, *}(M; L).
\]
\end{definition}
This operator is $G$-equivariant by equivariance of the connection $\nabla$ (assumption 6) and invariance of the almost complex structure $J$ (assumption 7).

The Dolbeault--Dirac operator defines an unbounded symmetric operator on the Hilbert space of $L^{2}$-sections
\[
H_M^L:= L^2(M, \Bigwedge^{0, *}T^*M\otimes L),
\]
with respect to the Liouville measure $dm$ on $M$. The metric on $\Bigwedge^{0, *}T^*M\otimes L$ comes from the given metric $H$ on $L$ and
from the Riemannian metric $B = \omega(\cdot, J \cdot)$
on $TM$. Because $M$ is metrically complete, the closure of $\DS_M^L$ is a self-adjoint operator on $H_M^L$.\footnote{\label{fn complete}This follows from the connection between Dirac
operators and Riemannian metrics as given for example in \cite{connes}, section VI.1, combined with Section 10.2 of \cite{HigsonRoe}. See also \cite{wolf} and  page 96 of \cite{friedrich}.}
So we can apply the functional calculus 
(see e.g.\ \cite{reedsimon}) to define the bounded operator
$F_M^L := b\bigl(\DS_M^L\bigr)$ on $H_M^L$, where $b$ is a \emph{normalising function} \cite{HigsonRoe}: 
\begin{definition} \label{def norm func}
A smooth function $b: \R \to \R$ is called a \emph{normalising function} if it has the following three properties:
\begin{itemize}
\item $b$ is odd;
\item $b(t) > 0$ for all $t > 0$;
\item $\lim_{t \to \pm \infty} b(t) = \pm 1$.
\end{itemize}
\end{definition}

Let $C_0(M)$ denote the $C^*$-algebra of continuous functions on $M$ that vanish at infinity, and let $\B(H_M^L)$ be the the $C^*$-algebra of bounded operators on 
$H_M^L$. Let
\[
\pi_M: C_0(M) \to \B(H_M^L)
\]
be the representation of $C_0(M)$ on $H_M^L$ defined by multiplication of sections by functions. Then the triple $(H_M^L, F_M^L, \pi_M)$ defines a $K$-homology class
\[
\bigl[\DS_M^L\bigr]:=[H_M^L, F_M^L, \pi_M] \in K_0^G(M),
\] 
which is independent of $b$.
See \cite{BCH, HigsonRoe, valette} for the definition of $K$-homology, and in particular Theorem 10.6.5 in \cite{HigsonRoe} for the claim that $\bigl[\DS_M^L\bigr]$ defines a $K$-homology class.

\begin{remark}
The Dolbeault--Dirac operator has the same principal symbol as the Spin$^c$ Dirac operator associated to the almost complex structure $J$ on $M$ and the line bundle $L$ 
(see e.g. \cite{duistermaat}, page 48), namely the Clifford action of $T^*M$ on $\Bigwedge^{0, *}T^*M \otimes L$. So the two operators define
the same class in $K$-homology. (The linear path  between the operators provides a homotopy.) Thus, if we consider the $K$-homology class $[\DS_M^L]$, we may take $\DS_M^L$ to
be either the Spin$^c$ Dirac operator or the Dolbeault--Dirac operator.
\end{remark}

\subsection{Reduction by $\Gamma$}

Because the subgroup $\Gamma$ of $G$ is discrete, the symplectic quotient 
 $M_{\Gamma}$
of $M$ by $\Gamma$ is equal to the orbit manifold $M/\Gamma$. The symplectic
form $\omega_{\Gamma}$ on $M/\Gamma$ is determined by
\[
p^*\omega_{\Gamma}=\omega,
\]
with $p: M \to M/\Gamma$ the quotient map.
The action of $G$ on $M$ descends to an action of $G/\Gamma$ on $M/\Gamma$, which satisfies the assumptions of Section \ref{sec group actions} (with $M$,
$\omega$, and $G$ replaced by $M/\Gamma$, $\omega_{\Gamma}$ and $G/\Gamma$, respectively).
The quotient $L/\Gamma$ turns out to be the total space of a prequantum line bundle over $M/\Gamma$. This is implied by the following fact.
\begin{proposition} \label{prop quotient vector bundle}
Let $H$ be a group acting properly and freely on a manifold $M$. Let $q: E \to M$ be an $H$-vector bundle. Then the induced projection
\[
q^H: E/H \to M/H
\]
defines a vector bundle over $M/H$.

Let $C^{\infty}(M, E)^H$ be the space of $H$-invariant sections of $E$. The linear map
\begin{equation} \label{eq psiE}
\psi_E: C^{\infty}(M, E)^H \to C^{\infty}(M/H, E/H),
\end{equation}
defined by
\[
\psi_E(s)(H\cdot m) = H \cdot s(m),
\]
is an \textit{isomorphism} of $C^{\infty}(M)^H \cong C^{\infty}(M/H)$-modules. 
\end{proposition}
Hence the quotient space $L/\Gamma$ is a complex line bundle over $M/\Gamma$, and its space of sections is
$C^{\infty}(M/\Gamma, L/\Gamma) \cong C^{\infty}(M, L)^{\Gamma}$.

The connection $\nabla$ on $L$ is $G$-equivariant, so it defines a $G/\Gamma$-equivariant connection $\nabla^{\Gamma}$ on $L/\Gamma$ as follows. Let 
$p: M \to M/\Gamma$
be the quotient map. Its tangent map
$Tp: TM \to T(M/\Gamma)$
induces an isomorphism of vector bundles over $M/\Gamma$
\[
Tp^{\Gamma}: (TM)/\Gamma \to T(M/\Gamma).
\]
Let
\[
\widetilde{Tp^{\Gamma}}: C^{\infty}(M/\Gamma, (TM)/\Gamma) \xrightarrow{\cong} \mathfrak{X}(M/\Gamma)
\]
be the isomorphism of $C^{\infty}(M/\Gamma)$ modules induced by $Tp^{\Gamma}$.
Consider the isomorphism of \mbox{$C^{\infty}(M)^{\Gamma} \cong C^{\infty}(M/\Gamma)$}-modules
\[
\varphi: \mathfrak{X}(M)^{\Gamma} \xrightarrow{\psi_{TM}} C^{\infty}(M/\Gamma, (TM)/\Gamma) \xrightarrow{\widetilde{Tp^{\Gamma}}} \mathfrak{X}(M/\Gamma).
\]
Let $v \in \mathfrak{X}(M)^{\Gamma}$ be a $\Gamma$-invariant vector field on $M$.  Then the operator $\nabla_v$ is $\Gamma$-equivariant, and hence maps $\Gamma$-invariant scetions of $L$ to invariant sections. The covariant derivative
$\nabla^{\Gamma}_{\varphi(v)}$ on $C^{\infty}(M/\Gamma, L/\Gamma)$ is defined by the commutativity of the following diagram:
\[
\xymatrix{C^{\infty}(M/\Gamma, L/\Gamma) \ar[r]^{\nabla^{\Gamma}_{\varphi(v)}} & C^{\infty}(M/\Gamma, L/\Gamma) \\
C^{\infty}(M, L)^{\Gamma} \ar[u]^{\psi_L}_{\cong} \ar[r]^{\nabla_v} & C^{\infty}(M, L)^{\Gamma}. \ar[u]^{\psi_L}_{\cong}}
\] 
A computation shows that $\nabla^{\Gamma}$ satisfies the properties of a connection, and that its curvature is
\[
\bigl(\nabla^{\Gamma}\bigr)^2=2\pi i \, \omega_{\Gamma}.
\]

Furthermore, the $G$-invariant Hermitian metric $H$ on $L$ descends to a $G/\Gamma$-invariant Hermitian metric $H^{\Gamma}$ 
on $L/\Gamma$, and the connection $\nabla^{\Gamma}$ is Hermitian with respect to this metric.
Finally, the $G$-invariant almost complex structure $J$ on $TM$ induces a $G/\Gamma$-invariant almost complex structure $J^{\Gamma}$ on $T(M/\Gamma)\cong (TM)/\Gamma$.
The corresponding Riemannian metric is denoted by
\[
B^{\Gamma} = \omega_{\Gamma}(\cdot, J^{\Gamma} \cdot ).
\]

From the Dirac operator $\DS_{\MGam}^{L/\Gamma}$ on $M/\Gamma$ associated to the almost complex structure $J^{\Gamma}$, coupled to the line bundle $L/\Gamma$ via $\nabla^{\Gamma}$, we form the bounded operator $F_{\MGam}^{L/\Gamma} := b\Bigl(\DS_{\MGam}^{L/\Gamma}\Bigr)$ (where $b$ is a normalising function) on the Hilbert space
\[
H_{M/\Gamma}^{L/\Gamma} := L^2\Bigl(\MGam, \Bigwedge^{0, *}T^*(\MGam)\otimes L/\Gamma \Bigr),
\]
which is defined with respect to the metrics on $\Bigwedge^{0, *}T^*(\MGam)$ and $L/\Gamma$ coming from those on $\Bigwedge^{0, *}T^*M$ and $L$ respectively,  and the measure $d\O$ on $\MGam$ defined as follows.

Let $U \subset M$ be a fundamental domain for the $\Gamma$-action. That is, $U$ is an open subset, $\Gamma \cdot U$ is dense in $M$, and if 
$m$ is a point in $U$, and $\gamma \in \Gamma$ is such that $\gamma \cdot m \in U$, then $\gamma = e$. Then for all measurable functions $f$ on $\MGam$ we define 
\begin{equation} \label{eq def measure dGamma m}
\int_{\MGam} f(\O) d\O := \int_U p^*f(m) dm, 
\end{equation}
where $p: M \to \MGam$ is the quotient map.
If $V$ is another fundamental domain, the subsets $\Gamma \cdot U$ and $\Gamma \cdot V$ differ by a set of measure zero, so this definition does not depend on the
choice of the fundamental domain. An equivalent way of defining $d\O$ is to say that the $d\O$-measure of a measurable subset $A \subset \MGam$ equals the
$dm$-measure of the subset $p^{-1}(A) \cap U$ of $M$. And since $dm$ is the Liouville measure on $(M, \omega)$, the measure $d\mathcal{O}$ is precisely the Liouville measure on $(M/\Gamma, \omega_{\Gamma})$.

We then have the $K$-homology class
\[
\Bigl[\DS_{\MGam}^{L/\Gamma}\Bigr] := [H_{\MGam}^{L/\Gamma}, F_{\MGam}^{L/\Gamma}, \pi_{\MGam} ] \in K_0^{\GGam}(\MGam).
\]

\subsection{The main result}

In Subsection \ref{subsec L^2} we define a homomorphism
\[
V_{\Gamma}: K_0^G(M) \to K_0^{\GGam}(\MGam),
\]
such that the following diagram commutes:
\begin{equation} \label{eq naturality mu}
\xymatrix{
K_0^G(M) \ar[r]^{\mu^G_M} \ar[d]^{V_{\Gamma}} & K_0(C^*(G)) \ar[d]^{R_Q^{(\Gamma)}}  \\
K_0^{G/\Gamma}(\MGam) \ar[r]^{\mu^{G/\Gamma}_{M/\Gamma}} & K_0(C^*(G/\Gamma)).
}
\end{equation}
Here $\mu^G_M$ and $\mu^{G/\Gamma}_{M/\Gamma}$ are analytic assembly maps (see \cite{BCH, Val1, valette}), and the homomorphism $R_Q^{(\Gamma)}$ is defined in \eqref{eq Gammared}.
In Appendix \ref{app nat}, we sketch how to generalise Valette's proof in \cite{valette} of commutativity of diagram \eqref{eq naturality mu} (`naturality of the assembly map') for discrete groups to the nondiscrete case.

The main step in our proof of Theorem \ref{thm GS} is the following:
\begin{theorem} \label{thm reduction}
The homomorphism $V_{\Gamma}$ maps the $K$-homology class of the Dirac operator $\DS_M^L$ to the $K$-homology class of the Dirac operator $\DS_{\MGam}^{L/\Gamma}$ on the
reduced space $\MGam$:
\[
V_{\Gamma}\left([\DS_M^L]\right) = \left[\DS_{M/\Gamma}^{L/\Gamma}\right] \in K_0^{G/\Gamma}(M/\Gamma).
\]
\end{theorem}
As we noted in the Introduction, Theorem \ref{thm GS} follows from Theorem \ref{thm reduction}, the naturality of the assembly map (diagram \eqref{eq naturality mu})  and the Guillemin--Sternberg conjecture for compact $G$ and $M$.

\begin{remark}
We will actually prove a stronger result than Theorem \ref{thm reduction}.  Write
\[
V_{\Gamma}\left([H_M^L, F_M^L, \pi_M]\right) = [H_{\Gamma}, F_{\Gamma}, \pi_{\Gamma}].
\]
Then there is a unitary isomorphism
\[
\chi: H_{\Gamma} \to H_{M/\Gamma}^{L/\Gamma}
\]
that intertwines the pertinent representations of $G/\Gamma$ and of $C(M/\Gamma)$, and the operators $F_{\Gamma}$ and $F_{M/\Gamma}^{L/\Gamma}$.
\end{remark}

\section{Differential operators on vector bundles} \label{sec diff ops}

In this section, we will compute the image under the homomorphism $V_{N}$ in diagram \eqref{eq naturality mu} 
of a $K$-homology class associated to an equivariant
elliptic first order differential operator on a vector bundle over a smooth manifold. The result is Corollary \ref{cor diff op}. 
In Section \ref{sec Dirac operators} we will see that Theorem \ref{thm reduction} is a special case of Corollary \ref{cor diff op}.

Let $G$ be a unimodular Lie group, and let $N$ be a closed normal subgroup of $G$. Let $dg$  and $dn$ be  Haar measures on $G$ and $N$ respectively. 
Let $M$ be a smooth manifold on which $G$ acts properly, such that the  action of $N$ on $M$ is free. Suppose $M/N$ is compact.\footnote{Compactness of $M/N$ 
is used in the proof of Lemma \ref{lem chi surjective}, but not in an essential way (see Footnote \ref{fn M/N cpt}).}

\subsection{The homomorphism $V_N$} \label{subsec L^2}
Let us briefly state the definition of the homomorphism
\[
V_N: K_0^G(M) \to K_0^{G/N}(M/N).
\]
For details we refer to  \cite{valette}. Let $H$ be a $\Z_2$-graded Hilbert space carrying a unitary representation of $G$, 
$F$ a $G$-equivariant bounded operator on $H$, and $\pi$ a representation of $C_0(M)$ in $H$ that is $G$-equivariant in the sense that for all $g \in G$ and $f \in C_0(M)$, one has $g \pi(f)g^{-1} = \pi(g \cdot f)$. Suppose $(H, F, \pi)$ defines a $K$-homology cycle. Then
\[
V_N[H, F, \pi] := [H_N, F_N, \pi_N],
\]
with $H_N$, $F_N$ and $\pi_N$ defined as follows.\footnote{The construction below originated in Rieffel's theory of induced representations of $C^*$-algebras \cite{Rie74},
which independently found its way into the \BCC\ \cite{BCH} and into the theory of constrained quantisation \cite{landsman95,landsman}.}

Consider the subspace $H_c := \pi(C_c(M))H \subset H$ and the sesquilinear  form $(\cdot, \cdot)_N$ on $H_c$ given by
\[
(\xi, \eta)_N := \int_N(\xi, n\cdot \eta)_H dn,
\]
for $\xi, \eta \in H_c$. This form turns out to be positive semidefinite. Consider the quotient space of $H_c$ by the kernel of this form, and complete this quotient in the
inner product $(\cdot, \cdot)_N$. This completion is $H_N$. 

Next, we use the fact that any $K$-homology class can be represented by a cycle whose operator is \emph{properly supported}:
\begin{definition} \label{def prop supp}
The operator $F$ is \emph{properly supported} if for every $f \in C_c(M)$ there is an $h \in C_c(M)$ such that $\pi(h)F\pi(f) = F \pi(h)$.
\end{definition}
Suppose $F$ is properly supported. Then it preserves $H_c$, and the restriction of $F$ to $H_c$ is bounded with respect to
the form $(\cdot, \cdot)_N$. Hence $F|_{H_c}$ induces a bounded operator $F_N$ on $H_N$ by continuous extension. The representation $\pi$ extends to the multiplier algebra
$C_b(M)$ of $C_0(M)$. The algebra $C_0(M/N)$ can be embedded into $C_b(M)$ via the isomorphism $C(M/N) \cong C(M)^N$, and then
we can use an argument similar to the one used in the definition of $F_N$ to show that $\pi$ induces a representation $\pi_N$ of $C_0(M/N)$ in $H_N$.

\subsection{Spaces of $L^2$-sections} \label{sec L2-spaces} 
Now let $q: E \to M$ be a $G$-vector bundle, equipped with a $G$-invariant metric $(\cdot, \cdot)_E$. Let $dm$ be a $G$-invariant measure on $M$, and let $L^2(M, E)$ be the
space of square-integrable sections of $E$ with respect to this measure. Let $\pi_M: C_0(M) \to \B(L^2(M, E))$ be the representation defined by multiplying sections by
functions. Let $L^2(M, E)_N$ be the Hilbert space constructed from $L^2(M, E)$ as in the definition of the homomorphism $V_N$. We will show that $L^2(M, E)_N$ 
is naturally isomorphic\footnote{A natural isomorphism between Hilbert spaces is an isomorphism defined without choosing bases of the spaces in question.}
to the Hilbert space $L^2(M/N, E/N)$ of square-integrable sections of the quotient vector bundle
\[
q_N: E/N \to M/N
\]
(see Proposition \ref{prop quotient vector bundle}). The $L^2$-inner product on sections of $E/N$ is defined via the
metric on $E/N$  induced by the one on $E$, and the measure $d\O$ on $M/N$  
 with the property
that for all measurable sections\footnote{Measurable in the sense that the inverse image of any Borel measurable subset of $M$ is Borel measurable in $M/N$.} 
$\varphi: M/N \to M$ and all $f \in C_c(M)$,
\begin{equation} \label{eq prop int}
\int_M f(m) dm = \int_{M/N} \int_N f(n \cdot \varphi(\O))\, dn\, d\O
\end{equation}
(see \cite{bourbaki}, Proposition 4b, p.\ 44). 
If $N$ is discrete and $dn$ is the counting measure, then the measure $d\O$ from \eqref{eq def measure dGamma m} satisfies property \eqref{eq prop int}.

Note that in this example, the space 
\[
L^2_c(M, E) := \pi(C_c(M))L^2(M, E)
\]
is the space of compactly supported $L^2$-sections of $E$.
Consider the linear map
\begin{equation} \label{eq def chi}
\chi: L^2_c(M, E) \to L^2(M/N, E/N),
\end{equation}
defined by
\[
\chi(s)(N  m) := N \cdot \int_N n \cdot s(n^{-1} m)\, dn,
\]
for all $s \in L^2_c(M, E)$ and $m \in M$. Because $s$ is compactly supported and the action is proper, the integrand is compactly supported for all $m \in M$. 
\begin{proposition} \label{prop chi iso}
The map $\chi$ induces a natural  $G/N$-equivariant unitary isomorphism
\begin{equation} \label{eq chi}
\chi:L^2(M, E)_{N} \xrightarrow{\cong} L^2(M/N, E/N).
\end{equation}
\end{proposition}
\begin{proof}
It follows from a lengthy but straightforward computation that the map $\chi$ is isometric, in the sense that for all $s \in L^2_c(M, E)$,
\[
\|\chi(s)\|_{L^2(M/N, E/N)} = \|s\|_{N},
\]
where $\|\cdot\|_N$ is the norm corresponding to the inner product $(\cdot, \cdot)_N$. Furthermore, $\chi$ is 
surjective, see Lemma \ref{lem chi surjective} below.
By these two properties, $\chi$ induces a bijective linear map
\begin{equation} \label{eq chi iso}
\chi: L^2_c(M, E)/ \mathcal{K} \to L^2(M/N, E/N),
\end{equation}
where $\mathcal{K}$ is the space of sections $s \in L^2_c(M, E)$ with $\|s\|_{N}=0$.
This map is a norm preserving linear isomorphism from $L^2_c(M, E)/\mathcal{K}$ onto the complete space $L^2(M/N, E/N)$. 
Hence the space $L^2_c(M, E)/\mathcal{K}$ is already complete, so that
$L^2(M, E)_{N} = L^2_c(M, E)/\mathcal{K}.$

So \eqref{eq chi iso} is actually a unitary isomorphism
\[
\chi: L^2(M, E)_{N} \to L^2(M/N, E/N).
\]
The fact that $N$ is a normal subgroup implies that this isomorphism intertwines the pertinent representations of $G/N$.
\end{proof}

\begin{lemma} \label{lem chi surjective}
The map $\chi$ in \eqref{eq def chi} is surjective.
\end{lemma}
\begin{proof}
Let $\sigma \in L^2(M/N, E/N)$. We will construct a section $s \in L^2_c(M, E)$ such that $\chi(s)=\sigma$, using the following diagram:
\[
\xymatrix{
E \ar[r]^{p_E} \ar[d]^{q} & E/N \ar[d]^{q_N} \\
M \ar[r]^{p} & M/N.
}
\]
Here the horizontal maps are quotient maps and define principal fibre bundles, and the vertical maps are vector bundle projections. 

Let $\{U_j\}$ be an open cover of $M/N$ that
admits local trivialisations
\[
\begin{split}
\tau_j:  p^{-1}(U_j) & \xrightarrow{\cong} U_j \times N \\
\theta^N_j: q_N^{-1}(U_j) & \xrightarrow{\cong} U_j \times E_0.
\end{split}
\]
Here $E_0$ is the typical fibre of $E$. Because $M/N$ is compact, the cover $\{U_j\}$ may be supposed to be finite.
Via the isomorphism of vector bundles $p^*(E/N)\cong E$, the trivialisations $\theta_j^N$ induce local trivialisations of $E$:
\[
\theta_j: q^{-1}(p^{-1}(U_j)) \xrightarrow{\cong} p^{-1}(U_j) \times E_0.
\]
And then, we can form trivialisations
\[
\tau_j^E: p_E^{-1}(q_N^{-1}(U_j)) \xrightarrow{\cong} q_N^{-1}(U_j) \times N,
\]
by
\[
\begin{split}
p_E^{-1}(q_N^{-1}(U_j)) &= q^{-1}(p^{-1}(U_j)) \\
	&\cong p^{-1}(U_j) \times E_0 \qquad \text{via $\theta_j$} \\
	&\cong U_j \times N \times E_0 \qquad \text{via $\tau_j$} \\
	&\cong q_N^{-1}(U_j) \times N \qquad \text{via $\theta_j^N$}.
\end{split}
\]
Here the symbol `$\cong$' indicates an $N$-equivariant diffeomorphism. It follows from the definition of the trivialisation $\theta_j$ that $\tau_j^E$ 
composed with projection onto $q_N^{-1}(U_j)$ equals $p_E$, so that $\tau_j^E$ is indeed an isomorphism of principal $N$-bundles.

For every $j$, define the section $s_j \in L^2(M, E)$ by
\[
s_j(\tau^{-1}_j(\O, n)) = \bigl(\tau_j^{E}\bigr)^{-1}(\sigma(\O), n)
\]
for all $\O \in U_j$ and $n \in N$, and extended by zero outside $p^{-1}(U_j)$.
By compactness\footnote{\label{fn M/N cpt} This is the only place where compactness of $M/N$ is used (the covering $\{U_j\}$ may also be locally finite). And even here, this assumption is not essential: it follows from this proof that all \emph{compactly supported} $L^2$-sections of $E/N$ are in the image of $\chi$. Hence $\chi$ has dense image, so that the induced map from $L^2(M, E)_N$ to $L^2(M/N, E/N)$ is surjective. By small adaptations to the proofs of Propositions \ref{prop chi iso} and \ref{prop chi D}, everything in this section still applies if $M/N$ is noncompact. But because we assume that the quotient $M/\Gamma$ is compact anyway, we will take the lazy option and suppose that $M/N$ is compact.}
of $M/N$, there is a compact subset $\widetilde{C} \subset M$ that intersects all $N$-orbits. Let $K \subset N$ be a compact subset of $dn$-volume $1$, and set 
$C := K \cdot \widetilde{C}$. Then for all $m \in M$, the volume of the compact set
\[
V_m := \{n \in N; n^{-1}m \in C\}
\] 
is at least $1$. Define the section $\tilde s$ of $E$ by
\[
\tilde s(m) = \left\{ \begin{array}{ll} \sum_j s_j(m) & \text{if $m \in C$} \\
	0 & \text{if $m \not\in C$.}
\end{array}\right.
\]
Then $\tilde s \in L^2_c(M,E)$, and for all $m \in M$,
\[
\begin{split}
\chi(\tilde s)(Nm) &= \sum_{\substack{j,\\ Nm \in U_j}} \int_{V_m} p_E \bigl(n\cdot s_j(n^{-1}m)\bigr) \, dn \\
	&= \sum_{\substack{j,\\ Nm \in U_j}} \int_{V_m} 
		p_E \bigl(\bigl(\tau_j^{E}\bigr)^{-1}(\sigma(Nm), n\cdot \psi_j(n^{-1}m) ) \bigr) \, dn,
\end{split}
\]
where $(Nm, \psi_j(n^{-1}m)) := \tau_j(n^{-1}m)$. Now since $p_E \circ \bigl(\tau_j^{E}\bigr)^{-1}$ is projection onto $q_N^{-1}(U_j)$, the latter integral equals
\[
\# \{j; Nm \in U_j\}\,  \mathrm{vol}(V_m) \, \sigma(Nm).
\]
Setting $\Phi(m):= \# \{j; Nm \in U_j\}\,  \mathrm{vol}(V_m)$ gives a measurable function $\Phi$ on $M$ which is bounded below by $1$ and $N$-invariant by invariance of $dn$. Hence
\[
s:= \frac{1}{\Phi}\tilde s,
\]
is a section 
$s \in L^2_c(M,E)$ for which $\chi(s)=\sigma$.
\end{proof}

\subsection{Differential operators} \label{sec diff op}

Let $G$ and $E \to M$ be as in Section \ref{sec L2-spaces}. Let
$D: C^{\infty}(M, E) \to C^{\infty}(M, E)$
be a $G$-equivariant first order differential operator that is symmetric with respect to the $L^2$-inner product on compactly supported sections.
Then $D$ defines an unbounded operator on $L^2(M, E)$. We assume that this operator has a self-adjoint extension, which we also denote by $D$.

\subsubsection*{Functional calculus and properly supported operators}
Applying the functional calculus to the self-adjoint extension of $D$, we define the bounded, self-adjoint operator $b(D)$ on
 $L^2(M, E)$, for any bounded measurable function $b$ on $\R$.\footnote{If $D$ is elliptic and $b$ is a normalising function , then $\bigl(L^2(M, E), b(D), \pi_M\bigr)$ is an equivariant $K$-homology cycle over $M$ (see Theorem 10.6.5 in \cite{HigsonRoe}).}
The operator $b(D)$ is $G$-equivariant because of the following result about functional calculus of unbounded operators, which follows directly from the definition as given for example in \cite{reedsimon}, page 261.
\begin{lemma} \label{prop func calc}
Let $H$ be a Hilbert space, and let $\mathcal{D} \subset H$ be a dense subspace. Let
$a:  \mathcal{D} \to H$
be a self-adjoint operator.
Let $H'$ be another Hilbert space, and let
$T: H \to H'$
be a unitary isomorphism.
Let $f$ be a measurable function on $\R$. Then
\[
Tf(a)T^{-1} = f(TaT^{-1}).
\]
\end{lemma}

We will later consider the case where $\bigl(L^2(M, E), b(D), \pi_M\bigr)$ is a $K$-homology cycle, and apply the map $V_N$ to this cycle. It is therefore important to us that the operator $b(D)$ is properly supported (Definition \ref{def prop supp}) for well-chosen functions $b$:
\begin{proposition} \label{prop norm prop supp}
If $b$ is a bounded measurable function with compactly supported (distributional) Fourier transform $\hat b$, then the operator $b(D)$ is properly supported.
\end{proposition}
The proof of this proposition is based on the following two facts, whose proofs can be found in \cite{HigsonRoe}, Section 10.3.
\begin{proposition} \label{prop eilD}
If $b$ is a bounded measurable function on $\R$ with compactly supported Fourier transform, then for all $s, t \in C^{\infty}_c(M, E)$,
\[
\bigl(b(D)s, t \bigr)_{L^2(M, E)} = \frac{1}{2 \pi}\int_{\R} \bigl(e^{i\lambda D}s, t \bigr)_{L^2(M, E)} \hat b(\lambda) \, d\lambda.
\]
\end{proposition} 
This is Proposition 10.3.5. from \cite{HigsonRoe}. By Stone's theorem, the operator $e^{i\lambda D}$ is characterised by the requirements that $\lambda \mapsto e^{i\lambda D}$ is a group homomorphism from $\R$ to the unitary operators on $L^2(M, E)$, and that for all 
$s \in C^{\infty}_c(M, E)$,
\[
\left. \frac{\partial}{\partial \lambda} \right|_{\lambda = 0}e^{i\lambda D}s = iDs.
\]
\begin{lemma} \label{lem eilD prop supp}
Let $s \in C^{\infty}_c(M, E)$, and let $h \in C^{\infty}_c(M)$ be equal to $1$ on the support of $s$. Let $\lambda \in \R$ such that
$|\lambda| < \|[D,\pi_M(h)]\|^{-1}$. Then
\[
\supp e^{i\lambda D}s \subset \supp h.
\]
\end{lemma}
This follows from the proof of Proposition 10.3.1. from \cite{HigsonRoe}.

\medskip
\noindent \emph{Proof of Proposition \ref{prop norm prop supp}.} Suppose $\supp \hat b \subset [-R, R]$. Let $f \in C_c(M)$, and choose $h \in C^{\infty}_c(M)$ such that $h$ equals $1$ on the support of $f$, and that $\|[D, \pi_M(h)]\| \leq \frac{1}{R}$. Let $1_M$ be the constant function $1$ on $M$. Then by 
Lemma \ref{lem eilD prop supp}, 
\begin{equation} \label{eq eilD}
\pi_M(1_M - h) e^{i\lambda D} \pi_M(f) = 0,
\end{equation}
for all $\lambda \in ]-R, R[$. Here we have extended the nondegenerate representation $\pi_M$ of $C_0(M)$ on $L^2(M, E)$ to the multiplier algebra $C_b(M)$ of $C_0(M)$. So by Proposition \ref{prop eilD}, we have for all $s, t \in C^{\infty}_c(M, E)$,
\[
\begin{split}
\bigl( \pi_M(1_M - h) b(D) \pi_M(f) s, t\bigr)_{L^2(M, E)} &= \bigl(  b(D) \pi_M(f) s, \pi_M(1_M - \bar h) t\bigr)_{L^2(M, E)}\\
	&= \frac{1}{2\pi} \int_{\R} \bigl(e^{i\lambda D} \pi_M(f) s, \pi_M(1_M - \bar h) t  \bigr)_{L^2(M, E)} \hat b(\lambda) \, d\lambda\\
	&= \frac{1}{2\pi} \int_{-R}^R \bigl(\pi_M(1_M - h) e^{i\lambda D} \pi_M(f) s,  t  \bigr)_{L^2(M, E)} \hat b(\lambda) \, d\lambda\\
	&=0,
\end{split}
\] 
by \eqref{eq eilD}. So
\[
\bigl(1-\pi_M(h)\bigr)b(D) \pi_M(f) = \pi_M(1_M - h) b(D) \pi_M(f) = 0,
\]
and hence $b(D)$ is properly supported.
\hfill $\blacksquare$

\subsubsection*{The image of $b(D)$ under $V_N$}
Now suppose that $D$ is elliptic and that $b$ is a normalising function with compactly supported Fourier transform,\footnote{If $g$ is a smooth, even, compactly supported function on $\R$, and $f := g * g$ is its convolution square, then $b(\lambda) := \int_{\R} \frac{e^{i\lambda x} - 1}{ix} f(x)\, dx$ is such a function (see \cite{HigsonRoe}, Exercise 10.9.3).} so that $b(D)$ is the kind of operator that defines a $K$-homology class over $M$. Because $b(D)$ is properly supported, it preserves $L^2_c(M, E)$ and the construction used in the definition of the map $V_N$ applies to $b(D)$. The resulting operator $b(D)_N$ on $L^2(M, E)_N$ is defined by commutativity of the following diagram:
\[
\xymatrix{L^2_c(M, E) \ar@{->>}[r] \ar[d]^{b(D)} & L^2(M, E)_N \ar[d]^{b(D)_N} \\
L^2_c(M, E) \ar@{->>}[r]  & L^2(M, E)_N.}
\]

On the other hand, the operator $D$ induces an unbounded operator on $L^2(M/N, E/N)$, because it restricts to
\[
\tilde D^N: C^{\infty}(M, E)^N \to C^{\infty}(M, E)^N,
\]
from which we obtain
\[
D^N := \psi_E^{-1}\tilde D^N \psi_E : C^{\infty}(M/N, E/N) \to C^{\infty}(M/N, E/N)
\]
(see Proposition \ref{prop quotient vector bundle}). We regard $D^N$ as an unbounded operator on $L^2(M/N, E/N)$. It is symmetric with respect to the $L^2$-inner product, and hence essentially self-adjoint by \cite{HigsonRoe}, Corollary 10.2.6. We therefore have the bounded operator $b(D^N)$ on $L^2(M/N,E/N)$.

Our claim is:
\begin{proposition} \label{prop chi D}
The isomorphism $\chi$ from Proposition \ref{prop chi iso} intertwines the operators $b(D)_N$ and $b(D^N)$:
\[
\xymatrix{
L^2(M, E)_N \ar[r]^-{\chi} \ar[d]^{b(D)_N} & L^2(M/N, E/N) \ar[d]^{b(D^N)} \\
L^2(M, E)_N \ar[r]^-{\chi}  & L^2(M/N, E/N).
}
\]
\end{proposition}
We will prove this claim by reducing it to the commutativity of another diagram. 
This diagram involves the Hilbert space $\tilde L^2(M/N, E/N)$, which is defined as the completion of the space $C^{\infty}(M, E)^N$ in the inner product
\[
(\sigma, \tau) := \int_{M/N} \bigl(\sigma(\varphi(\mathcal{O})), \tau(\varphi(\mathcal{O})\bigr)_E \, d\O, 
\]
for any measurable section $\varphi: M/N \to M$. The map $\psi_E$ from Proposition \ref{prop quotient vector bundle} extends continuously to a unitary isomorphism
\[
\tilde \psi_E: \tilde L^2(M/N, E/N)  \to  L^2(M/N, E/N).
\]

The unbounded operator $\tilde D^N$ on $\tilde L^2(M/N, E/N)$ is essentially self-adjoint because $D^N$ is, and because $\tilde \psi_E$ intertwines the two operators. Hence we have $b(\tilde D^N) \in \B\bigl(\tilde L^2(M/N, E/N)\bigr)$. We will deduce Proposition \ref{prop chi D} from
\begin{lemma} \label{lem chi D}
The following diagram commutes:
\[
\xymatrix{L^2_c(M, E) \ar[r]^-{\textstyle{\int_N} n\cdot} \ar[d]^{b(D)} & \tilde L^2(M/N, E/N) \ar[d]^{b(\tilde D^N)} \\
L^2_c(M, E) \ar[r]^-{\textstyle{\int_N} n\cdot}  & \tilde L^2(M/N, E/N),
}
\]
where the map $\textstyle{\int_N} n\cdot$ is given by\footnote{Note that the space $\tilde L^2(M/N, E/N)$ can be realised as a space of sections of $E$.}
\[
\Bigl({\textstyle{\int_N}} n\cdot(s)\Bigr)(Nm) = \int_N n\cdot s(n^{-1}m) \, dn.
\]
\end{lemma}
\begin{proof}

\noindent \emph{Step 1.} Because the representation of $N$ in $L^2(M, E)$ is unitary, we have
\[
\Bigl( \textstyle{\int_N} n \cdot (s), t\Bigr)_{L^2(M, E)} = 
	\Bigl(  s, \textstyle{\int_N} n \cdot (t)\Bigr)_{L^2(M, E)}
\]
for all $s, t \in L^2_c(M, E)$.

\medskip
\noindent \emph{Step 2.} By Proposition \ref{prop diff ops and int} below and equivariance of $D$, we have
\[
\Bigl(\textstyle{\int_N} n \cdot \Bigr)\circ D = \tilde D^N \circ \textstyle{\int_N} n \cdot
\]
on $C^{\infty}_c(M, E)$.

\medskip
\noindent \emph{Step 3.} For all $s \in C^{\infty}_c(M, E)$, we have by Proposition \ref{prop diff ops and int},
\[
\begin{split}
\left. \frac{\partial}{\partial \lambda} \right|_{\lambda = 0} \textstyle{\int_N} n \cdot \circ \, e^{i \lambda D} s &= 
		\int_N \frac{\partial}{\partial \lambda}  e^{i \lambda D} n \cdot  s\, dn \\
	&=i \int_N n \cdot Ds \, dn \\
	&= i \tilde D^N \textstyle{\int_N n\cdot} (s) \qquad \text{(by Step 2)} \\
	&= \left. \frac{\partial}{\partial \lambda} \right|_{\lambda = 0} e^{i \lambda \tilde D^N}\textstyle{\int_N} n \cdot  (s).
\end{split}
\]
So by Stone's theorem, 
\[
\textstyle{\int_N} n \cdot \circ \, e^{i \lambda D} = e^{i \lambda \tilde D^N} \circ \textstyle{\int_N} n \cdot
\]
for all $\lambda \in \R$.

\medskip
\noindent \emph{Step 4.} For all $s,t \in C^{\infty}_c(M, E)$,
\[
\begin{split}
\bigl( b(\tilde D^N) \textstyle{\int_N} n \cdot (s), t \bigr)_{L^2(M, E)} &= 
	\frac{1}{2 \pi}\int_{\R} \bigl(e^{i\lambda \tilde D^N} \textstyle{\int_N} n \cdot(s), t \bigr)_{L^2(M, E)} \hat b(\lambda) \, d\lambda 
			\qquad \text{(by Proposition \ref{prop eilD})}\\
	&= \frac{1}{2 \pi}\int_{\R} \bigl(\textstyle{\int_N} n \cdot e^{i\lambda D} s, t \bigr)_{L^2(M, E)} \hat b(\lambda) \, d\lambda 
			\qquad \text{(by Step 3)}\\
	&= \frac{1}{2 \pi}\int_{\R} \bigl( e^{i\lambda D} s, \textstyle{\int_N} n \cdot(t) \bigr)_{L^2(M, E)} \hat b(\lambda) \, d\lambda 
			\qquad \text{(by Step 1)}\\
	&= \bigl( b(\tilde D)  s, \textstyle{\int_N} n \cdot(t) \bigr)_{L^2(M, E)}
			\qquad \text{(by Proposition \ref{prop eilD})} \\
	&= \bigl( \textstyle{\int_N} n \cdot b(\tilde D)  s, t \bigr)_{L^2(M, E)}
			\qquad \text{(by Step 1).} \\
\end{split}
\]
This completes the proof.
\end{proof}
\begin{proposition} \label{prop diff ops and int}
Let $M_1$ and $M_2$ be manifolds, and suppose $M_2$ is equipped with a measure $dm_2$. Let $E \to M_1$ be a vector bundle, and let
\[
D: C^{\infty}(M_1, E)\to C^{\infty}(M_1, E)
\]
be a differential operator.  Let $p_1: M_1 \times M_2 \to M_1$ be projection onto the first factor, and consider the operator
\[
p_1^*D: C^{\infty}(M_1 \times M_2, p_1^*E) \to C^{\infty}(M_1 \times M_2, p_1^*E),
\]
defined by $\bigl(p_1^*D\bigr) p_1^*\sigma = p_1^* \bigl(D\sigma \bigr)$ for all $\sigma \in C^{\infty}(M_1, E)$.

Then for all $s \in C^{\infty}(M_1 \times M_2, p_1^*E)$ and all $m_1 \in M_1$,
\[
D\left(\int_{M_2} s(\cdot, m_2) dm_2\right)(m_1) = \int_{M_2} p_1^*D s(m_1, m_2) dm_2.
\]
\end{proposition}

We now derive Proposition \ref{prop chi D} from Lemma \ref{lem chi D}.

\medskip
\noindent \emph{Proof of Proposition \ref{prop chi D}.} Consider the following cube:
\[
\xymatrix{
L^2_c(M, E) \ar[r]^-{\textstyle{\int_N} n\cdot} \ar[d]_{b(D)} \ar@{->>}[ddr]& \tilde L^2(M/N, E/N) \ar[d]_{b(\tilde D^N)} \ar[ddr]^{\tilde \psi_E} & \\
L^2_c(M, E) \ar[r]|\hole \ar@{->>}[ddr] & \tilde L^2(M/N, E/N) \ar[ddr]|(0.51)\hole^(0.4){\tilde \psi_E} & \\
& L^2(M, E)_N \ar[r]_(0.4){\chi} \ar[d]^{b(D)_N} & L^2(M/N, E/N) \ar[d]^{b(D^N)} \\
& L^2(M, E)_N \ar[r]^-{\chi}  & L^2(M/N, E/N).
}
\]
The rear square (with the operators $b(D)$ and $b(\tilde D^N)$ in it) commutes by Lemma \ref{lem chi D}. The left hand square (with the operators $b(D)$ and $b(D)_N$) commutes by definition of $b(D)_N$, and the right hand square (with $b(\tilde D^N)$ and $b(D^N)$) commutes by Lemma \ref{prop func calc}. The top and bottom squares commute by definition of the map $\chi$, so that the front square commutes as well, which is Proposition \ref{prop chi D}.
\hfill $\blacksquare$

\subsection{Multiplication of sections by functions} \label{sec multiplication}
Let $G$, $M$ and $E$ be as in Subsections \ref{sec L2-spaces} and \ref{sec diff op}. As before, let
\[
\pi_M: C_0(M) \to \B(L^2(M, E))
\]
and 
\[
\pi_{M/N}: C_0(M/N) \to \B(L^2(M/N, E/N))
\]
be the representations defined by multiplication of sections by functions.
Let
\[
\left(\pi_M\right)_{N}: C_0(M/N) \to \B(L^2(M, E)_{N})
\]
be the representation obtained from $\pi_M$ by the procedure in Subsection \ref{subsec L^2}.
\begin{lemma} \label{lem piM}
The isomorphism \eqref{eq chi} intertwines the representations $\left(\pi_M\right)_{N}$ and $\pi_{M/N}$.
\end{lemma}
\begin{proof}
The representation $\left(\pi_M\right)_{N}$ is induced by
\begin{gather*}
\pi_M^{N}: C(M/N) \to \B(L^2_c(M, E)), \\
\left(\pi_M^{N}(f)s\right)(m) = f(N \cdot m)s(m).
\end{gather*}
For all $f \in C(M/N)$, $s \in L^2_c(M, E)$ and $m \in M$, we have
\[
\begin{split}
\chi\left(\pi_M^{N}(f)s\right)(N \cdot m) &= N \cdot \int_{N} n \cdot f(N \cdot n^{-1}m)s(n^{-1} \cdot m) \, dn\\
	&= N \cdot f(N \cdot m)  \int_{N} n \cdot s(n^{-1}\cdot m) \, dn \\
	&= \left( \pi_{M/N}(f)\chi(s)\right)(N \cdot m).
\end{split}
\]
\mbox{}
\end{proof}

\subsection{Conclusion} \label{sec conclusion}
Let $G$, $M$, $E$, $D$, $D^N$, $\pi_M$ and $\pi_{M/N}$ be as in Sections \ref{sec L2-spaces} -- \ref{sec multiplication}.
Suppose that the vector bundle $E$ carries a $\Z_2$-grading with respect to which the operator $D$
is odd. Suppose $D$ is elliptic and 
essentially self-adjoint as an unbounded 
 operator on $L^2(M, E)$.\footnote{This is the case if $M$ is complete and $D$ is a Dirac operator on $M$, see
footnote \ref{fn complete}.}. Let $b$ be a normalising function with compactly supported Fourier transform. Then Proposition \ref{prop chi iso}, 
Proposition \ref{prop chi D} and Lemma \ref{lem piM} may be summarised as follows.
\begin{theorem} \label{thm diff op}
Let $(L^2(M, E)_{N}, b(D)_{N}, \left(\pi_{M}\right)_{N})$ be the triple obtained from \\
$(L^2(M, E), b(D), \pi_{M})$ by the procedure of Subsection \ref{subsec L^2}. 
Then there is a unitary isomorphism
\[
\chi: L^2(M, E)_{N} \to L^2(M/N, E/N)
\]
that intertwines the representations of $G/N$, the operators $b(D)_{N}$ and $b(D^{N})$, 
and the representations $\left(\pi_{M}\right)_{N}$ and $\pi_{M/N}$.
\end{theorem}
\begin{corollary} \label{cor diff op}
The image of the class
\[
[D] := \Bigl[L^2(M, E), b(D), \pi_M \Bigr] \in K_0^{G}(M)
\]
under the homomorphism $V_N$ defined in Subsection \ref{subsec L^2} is
\[
V_N[D] = 
\Bigl[L^2(M/N, E/N), b(D^N), \pi_{M/N} \Bigr] =: \bigl[D^N \bigr]\in K_0^{G/N}(M/N). 
\]
\end{corollary}

\begin{remark} \label{rem non-noncomm GSC}
If the action of $G$ on $M$ happens to be free, then Corollary \ref{cor diff op} allows us to restate 
the Guillemin--Sternberg conjecture \ref{con [Q,R]=0}  without using techniques from 
noncommutative geometry. Indeed, for free actions
we have
\[
\begin{split}
R_Q \circ \mu^G_M \Bigl[ \DS_M^L \Bigr] &= \mu^{\{e\}}_{M/G} \circ V_G \Bigl[ \DS_M^L \Bigr] \qquad \text{by naturality of $\mu$} \\
	&= \ind \bigl(\DS_M^L\bigr)^G \qquad \text{by Corollary \ref{cor diff op}} \\
	&= \dim \Bigl(\ker \bigl( \DS_M^L\bigr)^+ \Bigr)^G - \dim \Bigl(\ker \bigl( \DS_M^L\bigr)^- \Bigr)^G \in \Z.
\end{split}
\]
Here the kernels of $\bigl(\DS_M^L\bigr)^{\pm}$ are taken in the spaces of smooth, not necessarily $L^2$, sections of $\Bigwedge^{0,*}T^*M \otimes L$.
Note that even though these kernels  may be infinite-dimensional, their $G$-invariant parts are not, because they are the kernels of the
elliptic operators $\Bigl(\bigl( \DS_M^L\bigr)^{\pm} \Bigl)^G$ on the compact manifold $M/G$.
So Conjecture \ref{con [Q,R]=0} becomes
\[
\ind \DS_{M_G}^{L_G} = \dim \Bigl(\ker \bigl( \DS_M^L\bigr)^+ \Bigr)^G - \dim \Bigl(\ker \bigl( \DS_M^L\bigr)^- \Bigr)^G.
\]
Unfortunately, in our situation this argument would only apply to discrete groups (see Remark \ref{rem discrete}).
\end{remark}

\section{Dirac operators} \label{sec Dirac operators}

In this section, we make the assumptions stated in Subsection \ref{sec group actions}. In particular, $\Gamma$ is a normal discrete subgroup of $G$. 
The goal of this section is to prove that Theorem \ref{thm reduction} is a special case of Corollary \ref{cor diff op}:
\begin{proposition} \label{prop Dirac operators}
Consider the Dolbeault--Dirac operator $\DS_M^L$ on $\Omega^{0,*}(M; L)$, and the induced operator $\bigl(\DS_M^L\bigr)^{\Gamma}$ 
on \mbox{$C^{\infty}\left(M/\Gamma, \bigl(\Bigwedge^{0, *} T^*M \otimes L\bigr)/\Gamma\right)$}. There is an isomorphism 
\[
\Xi: \Omega^{0,*}(M/\Gamma; L/\Gamma) \to C^{\infty}\left(M/\Gamma, \bigl(\Bigwedge^{0, *} T^*M \otimes L\bigr)/\Gamma \right)
\]
that is isometric with respect to the $L^2$-inner product and
intertwines the Dolbeault--Dirac operator $\DS_{M/\Gamma}^{L/\Gamma}$ on $\Omega^{0,*}(M/\Gamma; L/\Gamma) $ and the operator $\bigl(\DS_M^L\bigr)^{\Gamma}$.
\end{proposition}
Consequently, $\Xi$ induces a unitary isomorphism between the corresponding $L^2$-spaces, which by Lemma \ref{prop func calc} intertwines the bounded operators obtained 
from $\DS_{M/\Gamma}^{L/\Gamma}$ and $\bigl(\DS_M^L\bigr)^{\Gamma}$ using a normalising function with compactly supported Fourier transform. 
Hence Theorem \ref{thm reduction} follows, as
\[
\begin{split}
V_{\Gamma}\left(\bigl[\DS_M^L\bigr]\right) &= \left[\bigl(\DS_M^L\bigr)^{\Gamma}\right] \qquad \text{by Corollary \ref{cor diff op}} \\
	&= \bigl[\DS_{M/\Gamma}^{L/\Gamma}\bigr] \qquad \text{by Proposition \ref{prop Dirac operators}.}
\end{split}
\]

\subsection{The isomorphism}

The isomorphism of $C^{\infty}(M/\Gamma)$-modules $\Xi$ in Proposition \ref{prop Dirac operators} is defined as follows. The quotient map
$p: M \to M/\Gamma$
induces the vector bundle homomorphism
\begin{equation}  \label{eq Tp*}
\Bigwedge Tp^*:  \Bigwedge T^*(M/\Gamma) \to \Bigwedge T^*M. 
\end{equation}
Because $Tp$ intertwines the almost complex structures on $TM$ and $T(M/\Gamma)$, the homomorphism \eqref{eq Tp*} induces
\begin{equation} \label{eq Tp0,*}
\Bigwedge^{0, *} Tp^*: \Bigwedge^{0, *} T^*(M/\Gamma) \to \Bigwedge^{0,*}T^*M.
\end{equation}
Composition with the quotient map
$T^*M \to (T^*M)/\Gamma$
turns \eqref{eq Tp*} and \eqref{eq Tp0,*} into isomorphisms
\begin{align} 
\bigl(\Bigwedge Tp^*\bigr)^{\Gamma}:  \;  \Bigwedge T^*(M/\Gamma) & \to \bigl(\Bigwedge T^*M\bigr)/\Gamma;
    \label{eq Tp*Gamma} \\
\bigl(\Bigwedge^{0,*} Tp^*\bigr)^{\Gamma}:  \;  \Bigwedge^{0,*} T^*(M/\Gamma) & \to \bigl(\Bigwedge^{0,*} T^*M\bigr)/\Gamma. \label{eq Tp0,*Gamma}
\end{align}
On the spaces of smooth sections of the vector bundles in question, the isomorphisms \eqref{eq Tp*Gamma} and \eqref{eq Tp0,*Gamma} induce
isomorphisms of $C^{\infty}(M/\Gamma)$-modules
\begin{align} 
\Psi:  \Omega^*(M/\Gamma) & \to C^{\infty}\bigl(M/\Gamma, \left(\Bigwedge T^*M\right)/\Gamma \bigr); \label{eq iso psi} \\
\Psi^{0,*}:  \Omega^{0, *}(M/\Gamma) & \to C^{\infty}\left(M/\Gamma, \bigl(\Bigwedge^{0, *}T^*M\bigr)/\Gamma\right).
\end{align}
Now the isomorphism $\Xi$ is defined as
\begin{multline*}
\Xi: \Omega^{0, *}(M/\Gamma; L/\Gamma) \cong  \\
\Omega^{0, *}(M/\Gamma) \otimes_{C^{\infty}(M/\Gamma)} C^{\infty}(M/\Gamma, L/\Gamma) 
\xrightarrow{\Psi^{0,*} \otimes 1_{ C^{\infty}(M/\Gamma, L/\Gamma) }} \\
C^{\infty}\left(M/\Gamma, \bigl(\Bigwedge^{0,*}T^*M\bigr)/\Gamma \right) 
\otimes_{ C^{\infty}(M/\Gamma)} C^{\infty}(M/\Gamma, L/\Gamma) \\
\cong  C^{\infty}\left(M/\Gamma, \bigl(\Bigwedge^{0,*}T^*M \otimes L\bigr)/\Gamma \right) .
\end{multline*}
It is isometric by definition of the measure $d\O$ on $M/\Gamma$ and the metrics on the vector bundles involved. 
Therefore, it remains to prove that $\Xi$ intertwines the operators $\DS_{M/\Gamma}^{L/\Gamma}$ and $\bigl(\DS_M^L\bigr)^{\Gamma}$.

\subsection{Proof of Proposition \ref{prop Dirac operators}}

\subsubsection*{The connections}
Recall the isomorphism of \mbox{$C^{\infty}(M)^{\Gamma} \cong C^{\infty}(M/\Gamma)$}-modules $\psi_{E}: C^{\infty}(M, E)^{\Gamma} \to C^{\infty}(M/\Gamma, E/\Gamma)$ defined
by \eqref{eq psiE}, with $H=\Gamma$,
for any $\Gamma$-vector bundle $E$ over $M$. Also consider the pullback $p^*$ of differential forms on $M/\Gamma$ to invariant differential forms on $M$. 
It defines an isomorphism of \mbox{$C^{\infty}(M/\Gamma) \cong C^{\infty}(M)^{\Gamma}$}-modules
\[
p^*: \Omega^*(M/\Gamma) \to \Omega^*(M)^{\Gamma}.
\]
\begin{lemma} \label{prop connections}
The following diagram commutes:
\[
\xymatrix{
\Omega^*(M; L)^{\Gamma} \ar[r]^{\nabla} \ar[d]^{\cong}& \Omega^*(M; L)^{\Gamma} \ar[d]^{\cong} \\
\Omega^*(M)^{\Gamma} \otimes_{C^{\infty}(M)^{\Gamma}} C^{\infty}(M, L)^{\Gamma} & \Omega^*(M)^{\Gamma} \otimes_{C^{\infty}(M)^{\Gamma}} C^{\infty}(M, L)^{\Gamma} \\
\Omega^*(M/\Gamma) \otimes_{C^{\infty}(M/\Gamma)} C^{\infty}(M/\Gamma, L/\Gamma) \ar[u]^{p^*\otimes \psi_L^{-1}}_{\cong} & \Omega^*(M/\Gamma) \otimes_{C^{\infty}(M/\Gamma)} C^{\infty}(M/\Gamma, L/\Gamma) \ar[u]^{p^*\otimes \psi_L^{-1}}_{\cong}  \\
\Omega^*(M/\Gamma; L/\Gamma) \ar[r]^{\nabla^{\Gamma}} \ar[u]_{\cong}& \Omega^*(M/\Gamma; L/\Gamma) \ar[u]_{\cong}.
}
\]
\end{lemma}
The proof of this lemma is a matter of writing out definitions.

By definition of the almost complex structure on $T(M/\Gamma)$, we have 
\[
p^*\bigl(\Omega^{0, q}(M/\Gamma)\bigr) = \Omega^{0,q}(M)^{\Gamma}
\]
for all $q$. Therefore, Lemma \ref{prop connections} implies that the following diagram commutes:
\begin{equation} \label{diag connections complex}
\xymatrix{\Omega^{0,*}(M; L)^{\Gamma} \ar[r]^-{\bar \partial_L} & \Omega^{0,*}(M; L)^{\Gamma} \\
\Omega^{0,*}(M/\Gamma; L/\Gamma) \ar[r]^-{\bar \partial_L^{\Gamma}} \ar[u]^{p^* \otimes \psi_L^{-1}}_{\cong} & \Omega^{0,*}(M/\Gamma; L/\Gamma),
\ar[u]^{p^* \otimes \psi_L^{-1}}_{\cong}
} 
\end{equation}
with $\bar \partial_L$ and $\bar \partial_L^{\Gamma}$ as in Subsection \ref{sec Dirac oper}.

\subsubsection*{The Dirac operators}
By definition of the measure $d\O$ on $M/\Gamma$ and the metrics $B^{\Gamma}$ on $T(M/\Gamma)$ and $H^{\Gamma}$ on $L/\Gamma$, the isomorphism
\[
p^* \otimes \psi_L^{-1}: \Omega^{0,*}(M/\Gamma; L/\Gamma) \to \Omega^{0,*}(M; L)^{\Gamma}
\]
is isometric with respect to the inner product on \mbox{$\Omega^{0,*}(M/\Gamma; L/\Gamma) $} defined by 
\begin{equation} \label{eq inner prod 1}
(\alpha \otimes \sigma, \beta \otimes \tau) = \int_{M/\Gamma} B^{\Gamma}(\alpha, \beta)H^{\Gamma}(\sigma, \tau) d\O, 
\end{equation}
for all $\alpha, \beta \in \Omega^{0,*}(M/\Gamma)$ and $\sigma, \tau \in  C^{\infty}(M/\Gamma, L/\Gamma)$, and the inner product on $\Omega^{0,*}(M; L)^{\Gamma}$ defined by
\begin{equation} \label{eq inner prod 2}
(\zeta \otimes s, \xi \otimes t) = \int_U B(\zeta , \xi) H(s, t) dm,
\end{equation}
for all $\zeta, \xi \in \Omega^{0,*}(M)^{\Gamma}$ and $s, t \in C^{\infty}(M, L)^{\Gamma}$. (Recall that $U \subset M$ is a fundamental domain for the $\Gamma$-action.)

The Dolbeault--Dirac operators on $M$ and $M/\Gamma$ are defined by
\[
\begin{split}
\DS_M^L &= \bar \partial_L + \bar \partial_L ^*; \\
\DS_{M/\Gamma}^{L/\Gamma} &= \bar \partial_L^{\Gamma} + \left(\bar \partial_L^{\Gamma}\right)^*.
\end{split}
\]
Here the formal adjoint $\left(\bar \partial_L^{\Gamma}\right)^*$ is defined with respect to the inner product \eqref{eq inner prod 1}. 
The formal adjoint  $ \bar \partial_L ^*$ is defined by
\[
\int_M (B \otimes H)\left(\bar \partial_L ^* \eta, \theta \right) dm =  
\int_M (B \otimes H)\left(\eta, \bar \partial_L  \theta \right) dm,
\]
for all $\eta, \theta \in \Omega^{0,*}(M; L)$, $\theta$ with compact support. But this is actually the same as the formal adjoint 
of $\bar \partial_L $ with respect to the inner product \eqref{eq inner prod 2}:
\begin{lemma} \label{lem def adjoint}
Let $\Gamma$ be a discrete group, acting properly and freely on a manifold $M$, equipped with a $\Gamma$-invariant measure $dm$. Suppose $M/\Gamma$ is compact.
Let $E \to M$ be a $\Gamma$-vector bundle,
equipped with a $\Gamma$-invariant metric $\langle \cdot, \cdot \rangle$. Let
\[
D: C^{\infty}(M, E) \to C^{\infty}(M, E)
\]
be a $\Gamma$-equivariant differential operator. Let
\[
D^*: C^{\infty}(M, E) \to C^{\infty}(M, E)
\]
be the operator  
such that for all $s, t \in C^{\infty}(M, E)$, $t$ with compact support,
\[
\int_M\langle D^*s, t\rangle dm = \int_M\langle s, Dt\rangle dm.
\]

Let $U \subset M$ be a fundamental domain for the $\Gamma$-action.
Then the restriction of $D^*$ to $C^{\infty}(M, E)^{\Gamma}$ satisfies
\[
\int_U \langle D^*s, t \rangle dm = \int_U\langle s, Dt \rangle dm,
\]
for all $s, t \in C^{\infty}(M, E)^{\Gamma}$.
\end{lemma}
\begin{proof}
We will show that for all $s \in C^{\infty}(M, E)^{\Gamma}$, and all $t$ in a dense subspace of $C^{\infty}(M, E)^{\Gamma}$, we have
\[
\int_U \langle D^*s, t \rangle dm = \int_U\langle s, Dt \rangle dm.
\]

Let $\tau$ be a section of $E$, with compact support in $U$. Define the section $t$ of $E$ by extending the restriction $\tau|_U$ $\Gamma$-invariantly to $M$. The space of all
sections $t$ obtained in this way is dense in $C^{\infty}(M, E)^{\Gamma}$ with respect to the topology induced by the inner product
\[
(s, t) := \int_U \langle s, t \rangle dm.
\]

Then for all $s \in C^{\infty}(M, E)^{\Gamma}$,
\[
\int_U \langle D^*s, t \rangle dm = \int_M \langle D^*s, \tau \rangle dm 
	= \int_M \langle s, D\tau \rangle dm 
	= \int_U \langle s, Dt \rangle dm.
\]
\mbox{}
\end{proof}

We conclude that $p^* \otimes \psi_L^{-1}$ is an isometric isomorphism with respect to the inner products used to define the adjoints 
$ \bar \partial_L ^*$ and $ \left(\bar \partial_L^{\Gamma} \right)^* $. 
Hence the commutativity of diagram \eqref{diag connections complex} implies
\begin{corollary} \label{cor diagr dirac}
The following diagram commutes:
\[
\xymatrix{
\Omega^{0,*}(M; L)^{\Gamma} \ar[r]^{\DSS_M^L} & \Omega^{0,*}(M; L)^{\Gamma} \\
\Omega^{0,*}(M/\Gamma; L/\Gamma) \ar[r]^{\DSS_{M/\Gamma}^{L/\Gamma}} \ar[u]^{p^* \otimes \psi_L^{-1}}_{\cong} & \Omega^{0,*}(M/\Gamma; L/\Gamma) \ar[u]^{p^* \otimes 
\psi_L^{-1}}_{\cong}
.}
\]
\end{corollary}
\begin{remark}
Corollary \ref{cor diagr dirac} shows that for {\it discrete} groups a much stronger statement than the Guillemin-Sternberg conjecture holds. 
Indeed, by Remark \ref{rem non-noncomm GSC} the Guillemin-Sternberg
conjecture states that the restriction of the operator $\DSS_M^L$ to $\Omega^{0,*}(M; L)^{\Gamma}$ is related to the operator $\DSS_{M/\Gamma}^{L/\Gamma}$ by the fact that their indices are equal (as operators on smooth, not necessarily $L^2$, sections). 
But these operators are in fact more strongly related: they are intertwined by an isometric isomorphism.
\end{remark}

\subsubsection*{End of the proof of Proposition \ref{prop Dirac operators}}
The last step in the proof of Proposition \ref{prop Dirac operators} is a decomposition of the isomorphism
\[
p^*: \Omega^*(M/\Gamma) \to \Omega^*(M)^{\Gamma}.
\]
\begin{lemma} \label{lem deco p*}
The following diagram commutes:
\[
\xymatrix{
\Omega^*(M/\Gamma) \ar[r]^{p^*}_{\cong} \ar[d]_{\Psi}^{\cong} & \Omega^*(M)^{\Gamma} \ar[dl]^{\psi_{\wedge T^*M}}_{\cong} \\
C^{\infty}(M/\Gamma, ( \Bigwedge T^*M)/\Gamma), & 
}
\]
where $\Psi$ is the isomorphism \eqref{eq iso psi}.
\end{lemma}
The proof of this lemma is a short and straightforward computation.

\medskip

\noindent \textit{Proof of Proposition \ref{prop Dirac operators}.}
Together with Lemma \ref{lem deco p*} and the definition of the operator
\[
\bigl(\DS_M^L\bigr)^{\Gamma}: 
C^{\infty}\left(M/\Gamma, \bigl(\Bigwedge^{0,*}T^*M\otimes L\bigr)/\Gamma \right) 
		\to C^{\infty}\left(M/\Gamma, \bigl(\Bigwedge^{0,*}T^*M\otimes L\bigr)/\Gamma \right),
\]
Corollary \ref{cor diagr dirac}  implies that the following diagram commutes:
\[
\xymatrix{
\Omega^{0,*}(M; L)^{\Gamma} \ar[d]^{\cong}_{\psi_{\wedge^{0,*}T^*M} \otimes \psi_L} \ar[r]^{\DSS_M^L} & \Omega^{0,*}(M; L)^{\Gamma}  
	\ar[d]^{\cong}_{\psi_{\wedge^{0,*}T^*M} \otimes \psi_L} \\
C^{\infty}\left(M/\Gamma, \bigl(\Bigwedge^{0,*}T^*M \otimes L \bigr)/\Gamma \right) \ar[r]^{\bigl(\DSS_M^L\bigr)^{\Gamma}} &
	C^{\infty}\left(M/\Gamma, \bigl(\Bigwedge^{0,*}T^*M \otimes L \bigr)/\Gamma \right) \\
\Omega^{0,*}(M/\Gamma; L/\Gamma) \ar[u]_{\cong}^{\Xi = \Psi^{0,*}\otimes 1} \ar[r]^{\DSS_{M/\Gamma}^{L/\Gamma}} & \Omega^{0,*}(M/\Gamma; L/\Gamma). 
	\ar[u]_{\cong}^{\Xi = \Psi^{0,*}\otimes 1}
}
\]
Indeed, the outside diagram commutes by Corollary \ref{cor diagr dirac} and Lemma \ref{lem deco p*}, 
and the upper square commutes by definition of $\bigl(\DS_M^L\bigr)^{\Gamma}$. 
Hence the lower square commutes as well, which is Proposition \ref{prop Dirac operators}.
\hfill $\blacksquare$

\section{Abelian discrete groups} \label{Abeliancase}

In this section, we consider the situation of Section \ref{sec result}, with the additional assumption that $G=\Gamma$ is an abelian discrete group.
Then the Guillemin--Sternberg conjecture can be proved directly, without using naturality of the assembly map 
\eqref{eq naturality mu}. This proof is based on Proposition \ref{prop Dirac operators}, and the description of the assembly map in this special case given by Baum, Connes
and Higson \cite{BCH}, Example 3.11 (which in turn is based on Lusztig \cite{Lusztig}). We will first explain this example in a little more detail than given in \cite{BCH}, 
and then show how it implies Theorem \ref{thm GS} for abelian discrete groups.

\subsection{The assembly map for abelian discrete groups} \label{sec ass map abelian}

The proof of the Guillemin--Sternberg conjecture for discrete abelian groups is based on the following result:
\begin{proposition} \label{prop Valette ab}
Let $M$, $E$, $D$ and $D^{\Gamma}$ be as in  Subsection \ref{sec conclusion}. Suppose that
$G=\Gamma$ is abelian and discrete. Using the normalising function $b(x) = \frac{x}{\sqrt{1+x^2}}$, we form the operator
$F:= b(D)$, so that we have the class
\[
\left[L^2(M, E), F\right] \in K_0^{\Gamma}(M).
\]
Then\footnote{Recall that we abuse notation by writing $\ind D^{\Gamma} := \dim \ker \bigl(D^{\Gamma}\bigr)^+ - \dim \ker \bigl(D^{\Gamma}\bigr)^-$.}
\[
R^{(\Gamma)}_Q \circ \mu^{\Gamma}_M\left[L^2(M, E), F \right] = \ind D^{\Gamma}.
\]
\end{proposition}
In view of Proposition \ref{prop Dirac operators}, Proposition \ref{prop Valette ab} implies our Guillemin--Sternberg conjecture (i.e.\ Theorem \ref{thm GS}) 
for discrete abelian groups.

\subsubsection*{Kernels of operators as vector bundles}
Using Example 3.11 from  \cite{BCH}, we can explicitly compute
\begin{equation} \label{eq im mu}
[{\mathcal E}, F_{{\mathcal E}}] := \mu^{\Gamma}_M\left[L^2(M, E), F \right] \in K\! K_0(\C, C^*(\Gamma)).
\end{equation}
Note that since $\Gamma$ is discrete, its unitary dual $\hat \Gamma$ is compact. And because $\Gamma$ is abelian, all irreducible unitary representations are of the form
\[
U_{\alpha}: \Gamma \to U(1),
\]
for $\alpha \in \hat \Gamma$. Fourier transform defines an isomorphism $C^*(\Gamma) \cong C_0(\hat \Gamma)$. Therefore,
\[
K\! K_0(\C, C^*(\Gamma)) \cong K_0(C^*(\Gamma)) \cong K_0(C_0(\hat \Gamma)) \cong K^0(\hat \Gamma).
\]
Because $\hat \Gamma$ is compact, the image of $[{\mathcal E}, F_{{\mathcal E}}]$ in $K^0(\hat \Gamma)$ is the difference of the
isomorphism classes of two vector bundles over $\hat \Gamma$.
These two vector bundles can be determined as follows. 
For all $\alpha \in \hat \Gamma$, we define the Hilbert space $H_{\alpha}$ as the space of all measurable sections $s_{\alpha}$ of $E$ (modulo equality almost everywhere),
 such that for all $\gamma \in \Gamma$,
\[
\gamma \cdot s_{\alpha} = U_{\alpha}(\gamma)^{-1} s_{\alpha},
\]
and such that the norm
\begin{equation} \label{eq norm alpha}
\|s_{\alpha}\|_{\alpha}^2 = \langle s_{\alpha}, s_{\alpha}\rangle_{\alpha} 
\end{equation} 
is finite, where the inner product $\langle \cdot, \cdot \rangle_{\alpha}$ is defined by
\[
\langle s_{\alpha}, t_{\alpha}\rangle_{\alpha} := \int_{M/\Gamma} \bigl(s_{\alpha}(\varphi(\mathcal{O})), t_{\alpha}(\varphi(\mathcal{O}))\bigr)_E d\mathcal{O},
\]
where $\varphi$ is any measurable section of the principal fibre bundle $M \to M/\Gamma$. 
The space $H_{\alpha}$ is
isomorphic to the space of $L^2$-sections of the vector bundle $E_{\alpha}$, where
\[
E_{\alpha} := E/(\gamma \cdot e \sim U_{\alpha}^{-1}(\gamma)e) \to M/\Gamma.
\]

Let $H_{\alpha}^D$ be the dense subspace
\begin{equation}\label{eq H_alpha^D}
H_{\alpha}^D := \{s_{\alpha} \in H_{\alpha} \cap C^{\infty}(M, E); Ds_{\alpha} \in H_{\alpha}\} \subset H_{\alpha}.
\end{equation}
Because the operator $D$ is $\Gamma$-equivariant, it restricts to an unbounded operator
\[
D_{\alpha}: H_{\alpha}^D \to H_{\alpha}
\]
on $H_{\alpha}$. It is essentially self-adjoint by \cite{HigsonRoe}, Corollary 10.2.6., and hence induces the bounded operator
\begin{equation} \label{eq F alpha}
F_{\alpha}:= \frac{D_{\alpha}}{\sqrt{1+D_{\alpha}^2}} \in \mathcal{B}(H_{\alpha}).
\end{equation}
The grading on $E$ induces a grading on $H_{\alpha}$ with respect to which $D_{\alpha}$ and $F_{\alpha}$ are odd. The operators $F_{\alpha}$ are elliptic pseudo-differential operators:
\begin{lemma}
Let $D$ be an elliptic, first order differential operator on a vector bundle $E \to M$, and suppose $D$ defines an essentially self-adjoint operator on $L^2(M, E)$ with respect to some measure on $M$ and metric on $E$. Then the operator $F := \frac{D}{\sqrt{1+D^2}}$ is an elliptic pseudo-differential operator.
\end{lemma}
\begin{proof}
This result seems well known and is easily derived from a  
number of related results in the literature. It is, in any case,  
sufficient to show that  $(1+D^2)^{-\frac{1}{2}}$ is a pseudo- 
differential operator.
The following proof was communicated to us by Elmar Schrohe.\footnote 
{An independent proof was suggested to us by John Roe, who mentioned  
that in the case at hand the functional calculus for (pseudo) 
differential operators developed in \cite{Taylor} for compact  
manifolds may be extended to the noncompact case. A third proof may  
be constructed using heat kernel techniques, as in the unpublished  
Diplomarbeit of Hanno Sahlmann (Rainer Verch, private communication).}

According to \cite{Beals}, a bounded operator $A:L^2(\R^n) \rightarrow  
L^2(\R^n)$
is a pseudo-differential operator on $\R^n$ iff all iterated  
commutators with $x_j$
(as a multiplication operator)  and $\partial_{x_j}$ are bounded  
operators. This immediately yields
the lemma for $M=\R^n$ (cf.\ \cite{Beals}, Theorem 4.2). To extend  
this result to the manifold case,
we recall that  an operator $A:C^{\infty}(M)\to \mathcal{D}'(M)$ on  
a manifold $M$ is a pseudo-differential operator when for each choice  
of smooth functions $f$, $g$ with support in a single coordinate   
neighbourhood, $fAg$  is  a pseudo-differential operator on
$\R^n$. (Here one has to admit nonconnected coordinate neighbourhoods.)

Now write $(1+D^2)^{-\frac{1}{2}}$ as a Dunford integral (cf.\ \cite 
{DS1}, pp.\ 556--577), as follows:
$$ (1+D^2)^{-\frac{1}{2}}= \oint_C \frac{dz}{2\pi i}\, (1+z)^{-\frac 
{1}{2}}(z-D^2)^{-1}. $$
To compute the  commutators of $f(1+D^2)^{-\frac{1}{2}}g$ with  $x_j$  and $\partial_{x_j} 
$, one may  take these inside the contour integral. Boundedness of  
all iterated commutators then easily follows, using the fact that $f$  
and $g$ have compact support.

The same argument, with the exponent $-\frac{1}{2}$ replaced by $\frac{1}{2}$, shows that  $(1+D^2)^{\frac{1}{2}}$ is  
a pseudo-differential operator, and ellipticity of $(1+D^2)^{-\frac{1}{2}}$ follows.
\end{proof}

Consider the field of Hilbert spaces
\[
(H_{\alpha})_{\alpha \in \hat \Gamma}\to \hat \Gamma.
\]
In the next subsection, we will give this field the structure of a \textit{continuous} field of Hilbert spaces by specifying its space of
continuous sections $C(\hat \Gamma, (H_{\alpha})_{\alpha \in \hat \Gamma})$.  Consider the subfields
\[
\begin{split}
\left(\ker D^+_{\alpha}\right)_{\alpha \in \hat \Gamma}  & \to \hat \Gamma; \\
\left(\ker D^-_{\alpha}\right)_{\alpha \in \hat \Gamma}  & \to \hat \Gamma.
\end{split}
\]
These are indeed well-defined subfields of $(H_{\alpha})_{\alpha \in \hat \Gamma}$ because $\ker D^{\pm}_{\alpha} = \ker F^{\pm}_{\alpha}$
by the elliptic regularity theorem.

Suppose that $\left(\ker D_{\alpha}\right)_{\alpha \in \hat \Gamma}$ and $\left(\coker D_{\alpha}\right)_{\alpha \in \hat \Gamma}$ are vector bundles over $\hat \Gamma$ in
the relative topology. As in the proof of the Atiyah--J\"{a}nich Theorem (cf.\ \cite{WO}), the operator $D$ can always be replaced by another operator in such a way 
that the class $[L^2(M, E), F]$ does not change, and that 
these `fields of vector spaces' are indeed vector bundles (see also \cite{Laf1}). Then:
\begin{proposition} \label{prop im mu}
The image of the class $\left[L^2(M, E), F \right] \in K_0^{\Gamma}(M)$ under the assembly map $\mu^{\Gamma}_M$ is
\[
\mu^{\Gamma}_M\left[L^2(M, E), F \right] = \left[\left(\ker D^+_{\alpha}\right)_{\alpha \in \hat \Gamma}\right] 
- \left[\left(\ker D^-_{\alpha}\right)_{\alpha \in \hat \Gamma}\right]  \in K^0(\hat \Gamma).
\]
\end{proposition}
Proposition \ref{prop im mu} wil be proved in the next two subsections.

\subsection{The Hilbert $C^*$-module part of the assembly map}
In this subsection we determine the Hilbert $C^*(\Gamma)\cong C_0(\hat \Gamma)$-module ${\mathcal E}$ in \eqref{eq im mu} (Proposition \ref{prop module E}). 

\subsubsection*{A unitary isomorphism}
Let $d\alpha$ be the measure on $\hat \Gamma$ corresponding to the counting measure on $\Gamma$ via the Fourier transform. Consider the Hilbert space 
\[
H := \int_{\hat \Gamma}^{\oplus} H_{\alpha} d\alpha.
\]
That is, $H$ consists of the measurable maps
\[
s: \hat \Gamma \to (H_{\alpha})_{\alpha \in \hat \Gamma}, \qquad \alpha \mapsto s_{\alpha},
\]
such that $s_{\alpha} \in H_{\alpha}$ for all $\alpha$, and
\[
\|s\|^2_H = \langle s, s \rangle_H := \int_{\hat \Gamma} \|s_{\alpha}\|_{\alpha}^2 d\alpha < \infty.
\]

Define the linear map
$V: H \to L^2(M, E)$
by
\[
\left(Vs\right)(m) := \int_{\hat \Gamma} s_{\alpha}(m) d\alpha.
\]
\begin{lemma} \label{lem V iso}
The map $V$ is a unitary isomorphism, with inverse 
\begin{equation} \label{eq V-1}
\bigl(V^{-1}\sigma\bigr)_{\alpha}(m) = \sum_{\gamma \in \Gamma} \gamma \cdot \sigma(\gamma^{-1}m) U_{\alpha}(\gamma),
\end{equation}
for all $\sigma \in C_c(M, E) \subset L^2(M, E)$.
\end{lemma}
\begin{remark}
It follows from unitarity of $V$ that $Vs$ is indeed an $L^2$-section of $E$ for all $s \in H$. Conversely, a direct computation shows that for all 
$\sigma \in L^{2}(M, E)$, $\alpha \in \hat \Gamma$ and $\gamma \in \Gamma$, one has
\[
\gamma \cdot \bigl(V^{-1}\sigma\bigr)_{\alpha} = U_{\alpha}(\gamma)^{-1}\bigl(V^{-1}\sigma\bigr)_{\alpha},
\]
so that $V^{-1}\sigma$ lies in $H.$
\end{remark}
\textit{Sketch of proof of Lemma \ref{lem V iso}.}
The proof is based on the observations that for all $\alpha \in \hat \Gamma$,
\begin{equation} \label{eq delta alpha}
\sum_{\gamma \in \Gamma} U_{\alpha}(\gamma) = \delta_{1}(\alpha),
\end{equation}
where $\delta_1 \in \mathcal{D}'(\hat \Gamma)$ is the $\delta$-distribution at the trivial representation $1 \in \hat \Gamma$, and that for all $\gamma \in \Gamma$,
\begin{equation} \label{eq delta gamma}
\int_{\hat \Gamma} U_{\alpha}(\gamma) d\alpha = \delta_{\gamma e},
\end{equation}
the Kronecker delta of $\gamma$ and the identity element.
Using these facts, one can easily verify that $V$ is an isometry and that \eqref{eq V-1} is indeed the inverse of $V$.
\hfill $\blacksquare$
\medskip

The representation $\pi_H$ of $\Gamma$ in $H$ corresponding to the standard representation of $\Gamma$ in $L^2(M, E)$ via the isomorphism $V$ is given by
\[
\left(\pi_H(\gamma)s\right)_{\alpha} = U_{\alpha}(\gamma)^{-1}s_{\alpha}.
\]
This follows directly from the definitions of the space $H_{\alpha}$ and the map $V$.

\subsubsection*{Fourier transform}
The Hilbert $C^*(\Gamma)$-module ${\mathcal E}$ is the closure of the space $C_c(M, E)$ in the norm
\[
\|\sigma\|_{{\mathcal E}}^2 := \|\gamma \mapsto \langle \sigma, \gamma \cdot \sigma \rangle_{L^2(M, E)} \|_{C^*(\Gamma)}.
\]
The $C^*(\Gamma)$-module structure of ${\mathcal E}$ is defined by
\[
f \cdot \sigma = \sum_{\gamma \in \Gamma} f(\gamma) \, \gamma \cdot \sigma,
\]
for all $f \in C_c(\Gamma)$ and $\sigma \in C_c(M, E)$.
The isomorphism $V$ induces an isomorphism of the Hilbert $C^{*}(\Gamma)$-module ${\mathcal E}$ with the closure ${\mathcal E}_{H}$ of
$V^{-1}(C_c(M, E)) \subset H$
in the norm
\[
\|s\|_{{\mathcal E}_H}^2 := \|\gamma \mapsto \langle Vs, \gamma \cdot Vs \rangle_{L^2(M, E)}\|_{C^*(\Gamma)} 
	= \|\gamma \mapsto \langle s, \pi_H(\gamma) s \rangle_H \|_{C^*(\Gamma)},
\]
by unitarity of $V$.
The $C^*(\Gamma)$-module structure on ${\mathcal E}_H$ corresponding to the one on ${\mathcal E}$ via $V$ is given by
\begin{equation} \label{eq rep in E_H}
f\cdot s = \sum_{\gamma \in \Gamma} f(\gamma) \pi_H(\gamma) s,
\end{equation}
for all $f \in C_c(\Gamma)$ and $s \in V^{-1}(C_c(M, E))$.

Next, we use the isomorphism
$C_0(\hat \Gamma) \cong C^{*}(\Gamma) $
defined by the Fourier transform $\psi \mapsto \hat \psi$,
where
\[
\hat \psi(\gamma) = \int_{\hat \Gamma} \psi(\alpha)U_{\alpha}(\gamma) d\alpha
\]
for all $\psi \in C_c(\hat \Gamma)$.
Because of \eqref{eq delta alpha} and \eqref{eq delta gamma}, the inverse Fourier transform is given by
$ f \mapsto \hat f$, where for $f \in C_c(\Gamma)$, one has
\[
\hat f(\alpha) = \sum_{\gamma \in \Gamma} f(\gamma) U_{\alpha}(\gamma)^{-1}.
\]

So by Fourier transform, the Hilbert $C^*(\Gamma)$-module ${\mathcal E}_H$ corresponds to the Hilbert $C_0(\hat \Gamma)$-module $\hat {\mathcal E}_H$, which is the closure of
the space $V^{-1}(C_c(M, E))$ in the norm
\begin{align}
\|s\|_{\hat {\mathcal E}_H}^2 &= \Bigl\| \alpha \mapsto \sum_{\gamma \in \Gamma} \langle s, \pi_H(\gamma)s\rangle_H U_{\alpha}(\gamma)^{-1} \Bigr\|_{C_0(\hat \Gamma)} \nonumber \\
	&= \sup_{\alpha \in \hat \Gamma} \, \Bigl| \sum_{\gamma \in \Gamma} \langle s, \pi_H(\gamma)s\rangle_H U_{\alpha}(\gamma)^{-1}\Bigr|. \label{eq norm hat E}
\end{align}

\subsubsection*{Continuous sections}
Using the following lemma, we will describe the Hilbert $C_0(\hat \Gamma)$-module $\hat {\mathcal E}_H$ as the space of continuous sections of a continuous field of Hilbert spaces.
\begin{lemma} \label{lem norm hat E}
For all $s, t \in V^{-1}(C_c(M, E))$,
\[
\sum_{\gamma \in \Gamma} \langle s, \pi_H(\gamma)t\rangle_H U_{\alpha}(\gamma)^{-1} = \langle s_{\alpha}, t_{\alpha}\rangle_{\alpha}.
\]
\end{lemma}
\begin{proof}
Let $\varphi$ be a measurable section of the principal fibre bundle $M \to M/\Gamma$. Then
\[
\begin{split}
 \sum_{\gamma \in \Gamma} \langle s, \pi_H(\gamma)t\rangle_H U_{\alpha}(\gamma)^{-1} &=   
 \sum_{\gamma \in \Gamma} \Bigl(\int_{\hat \Gamma} \int_{M/\Gamma} \bigl(s_{\beta}(\varphi(\mathcal{O})), U_{\beta}(\gamma)^{-1}t_{\beta}(\varphi(\mathcal{O}))\bigr)_E 
							d\mathcal{O} \, d\beta \Bigr) U_{\alpha}(\gamma)^{-1}  \\
 &= \int_{M/\Gamma} \bigl(s_{\alpha}(\varphi(\mathcal{O})), t_{\alpha}(\varphi(\mathcal{O}))\bigr)_E d\mathcal{O}  \\
 &= \langle s_{\alpha}, t_{\alpha}\rangle_{\alpha}. 
\end{split}
\]
by \eqref{eq delta alpha}.
\end{proof}

We conclude from \eqref{eq norm hat E} and Lemma \ref{lem norm hat E} that $\hat {\mathcal E}_H$ is the closure of $V^{-1}(C_c(M, E))$ in the norm
\[
\|s\|_{\hat {\mathcal E}_H}^2 = \sup_{\alpha \in \hat \Gamma} \|s_{\alpha}\|_{\alpha}^{2}.
\]
Therefore, it makes sense to \textit{define} the space $C(\hat \Gamma, \bigl( H_{\alpha} \bigr)_{\alpha \in \hat \Gamma})$ of continuous 
sections of the field of Hilbert spaces
$(H_{\alpha})_{\alpha \in \hat \Gamma}$ as the $C_0(\hat \Gamma)$-module $\hat {\mathcal E}_H$
(cf.\ \cite{dixmierdouady,Taka}). 
Then our construction implies 
\begin{proposition} \label{prop module E}
The Hilbert $C^{*}(\Gamma)$-module ${\mathcal E}$ is isomorphic to the Hilbert $C_0(\hat \Gamma)$-module 
\mbox{$C(\hat \Gamma, (H_{\alpha})_{\alpha \in \hat \Gamma})$}.
\end{proposition}
Let us verify explicitly that the representations of $C_0(\hat \Gamma)$ in $\hat {\mathcal E}_H$ and in 
\mbox{$C(\hat \Gamma, (H_{\alpha})_{\alpha \in \hat \Gamma})$} are indeed 
intertwined by the isomorphism induced by $V$ and the Fourier transform:
for all $f \in C_c(\Gamma)$ and all $s \in V^{-1}(C_c(M, E))$, we have
\[
\begin{split}
( f \cdot s )_{\alpha} &= \sum_{\gamma \in \Gamma} f(\gamma) \left(\pi_{H}(\gamma)s\right)_{\alpha} \qquad\text{by \eqref{eq rep in E_H}}\\
	&= \sum_{\gamma \in \Gamma} f(\gamma) U_{\alpha}(\gamma)^{-1} s_{\alpha} \\
	&= \hat f(\alpha)s_{\alpha}.
\end{split}
\]

\subsection{The operator part of the assembly map}

\begin{proposition} \label{prop operator ass}
The operator $F_{\hat {\mathcal E}_H}$ on the Hilbert $C_0(\hat \Gamma)$ module 
$\hat {\mathcal E}_H = C(\hat \Gamma, (H_{\alpha})_{\alpha \in \hat \Gamma})$,
 induced by $F \in \mathcal{B}(L^2(M, E))$, equals
$F_{\hat {\mathcal E}_H} = \bigl(F_{\alpha}\bigr)_{\alpha \in \hat \Gamma}$,
where the `field of operators' $\bigl(F_{\alpha}\bigr)_{\alpha \in \hat \Gamma}$ is given by
\[
\left(\bigl(F_{\alpha}\bigr)_{\alpha \in \hat \Gamma} s\right)_{\beta} = F_{\beta}s_{\beta},
\]
for all $\beta \in \hat \Gamma$ and all $s \in V^{-1}(C_c(M, E))$ (and extended continuously to general $s \in \hat {\mathcal E}_H$). 
Here $F_{\alpha}$ is the operator \eqref{eq F alpha}.
\end{proposition}
\begin{proof}
For $s \in V^{-1}(C_c(M, E))$, we have
$V F_{\hat {\mathcal E}_H}s=F Vs$.
So it is sufficient to prove that for such $s$, one has
$FVs(m) = \int_{\hat \Gamma} F_{\alpha}s_{\alpha}(m) d\alpha$,
for all $m \in M$.

Let $H^D \subset H$ be the space of $s \in H$ such that $Vs \in C^{\infty}_c(M, E)$, and $s_{\alpha} \in H_{\alpha}^D$ for all $\alpha \in \hat \Gamma$ 
(see \eqref{eq H_alpha^D}).
By Proposition \ref{prop diff ops and int}, we have  
$DVs(m) = \int_{\hat \Gamma} Ds_{\alpha}(m) d\alpha$
for all $s \in H^D$ and $m \in M$.  Because of  Lemma \ref{prop func calc}  this proves the
proposition, because $H^D$ is dense in $H$.

Note that $\bigl(F_{\alpha}\bigr)_{\alpha \in \hat \Gamma}$ is initially defined on the subspace $V^{-1}(C_c(M, E))$ of 
$C(\hat \Gamma, (H_{\alpha})_{\alpha \in \hat \Gamma})$. But since the unitary operator $V$ intertwines $\bigl(F_{\alpha}\bigr)_{\alpha \in \hat \Gamma}$
and the bounded operator $F$ on $L^2(M, E)$, the field of operators $\bigl(F_{\alpha}\bigr)_{\alpha \in \hat \Gamma}$ is bounded in the norm $\|\cdot\|_{\hat\E_H}$, 
so that it extends continuously to a bounded operator on $C(\hat \Gamma, (H_{\alpha})_{\alpha \in \hat \Gamma})$.
\end{proof}

\textit{Proof of Proposition \ref{prop im mu}.}
We have seen (cf. Propositions \ref{prop module E} and \ref{prop operator ass}) that
\[
\mu^{\Gamma}_M\left[L^2(M, E), F\right] = \biggl[C(\hat \Gamma, (H_{\alpha})_{\alpha \in \hat \Gamma}), \left(F_{\alpha}\right)_{\alpha \in \hat \Gamma}\biggr]
	\in K\! K_0(\C, C_0(\hat \Gamma)).
\]
The image of this class in $K_0(C_0(\hat{\Gamma}))$ is the formal difference of projective
$C_0(\hat{\Gamma})$-modules 
\begin{equation} \label{eq kernels F}
\left[\ker \left(\left(F^+_{\alpha}\right)_{\alpha \in \hat \Gamma}\right) \right] - \left[\ker \left(\left(F^-_{\alpha}\right)_{\alpha \in \hat \Gamma}\right) \right].
\end{equation}
By compactness of $M/\Gamma$ and the elliptic regularity theorem, 
the kernels of $F^+_{\alpha}$ and $F^-_{\alpha}$ are equal to the kernels of $D^+_{\alpha}$ and $D^-_{\alpha}$,
respectively. If we suppose that the kernels of  $D^+_{\alpha}$ and $D^-_{\alpha}$ define vector bundles over $\hat \Gamma$, then by Lemma \ref{lem Bostop} below,
the class \eqref{eq kernels F} equals
\[
\left[C(\hat \Gamma, \left(\ker D^+_{\alpha}\right)_{\alpha \in \hat \Gamma} )\right] - 
	\left[C(\hat \Gamma, \left(\ker D^-_{\alpha}\right)_{\alpha \in \hat \Gamma} )\right].
\]

Under the isomorphism $K_0(C_0(\hat \Gamma)) \cong K^0(\hat \Gamma)$, the latter class corresponds to 
\[
\left[\left(\ker D^+_{\alpha}\right)_{\alpha \in \hat \Gamma} \right] - \left[\left(\ker D^-_{\alpha}\right)_{\alpha \in \hat \Gamma} \right] \in K^0(\hat \Gamma).
\]
\hfill $\blacksquare$
\begin{lemma} \label{lem Bostop}
Let $\mathcal{H}$ be a continuous field of Hilbert spaces over a topological space $X$, and let $\Delta$ be its space of continuous sections. Let $\mathcal{H}'$ be a
subset of $\mathcal{H}$ such that for all $x \in X$, $\mathcal{H}'_x := \mathcal{H}_x \cap \mathcal{H}' $ is a linear subspace of $\mathcal{H}_x$. Set
\[
\Delta' := \{s \in \Delta;  s(x) \in \mathcal{H}'_x \text{ for all } x \in X\}.
\] 

Let $s: X \to \mathcal{H}'$ be a section. Then $s$ is continuous in the subspace topology of $\mathcal{H}'$ in $\mathcal{H}$ if and only if $s \in \Delta'$.
\end{lemma}
\begin{proof}
Let $s: X \to \mathcal{H}$ be a section. Then $s$ is a continuous section of $\mathcal{H}'$ in the subspace topology if and only if $s$ is a continuous section of 
$\mathcal{H}$ and $s(x) \in \mathcal{H}'_x$ for all $x$. The topology on $\mathcal{H}$ is defined in such a way that $s$ is continuous if and only if $s \in \Delta$ 
\cite{dixmierdouady,Taka}.
\end{proof}

\subsection{Reduction}

We will now describe the reduction map
$R^{(\Gamma)}_Q: K_0(C^*(\Gamma)) \to \Z,$
and prove Proposition \ref{prop Valette ab}.

\begin{lemma} \label{lem reduction}
Let $\Gamma$ be an abelian discrete group, and let
$i: \{1\} \hookrightarrow \hat \Gamma$
be the inclusion of the trivial representation.
The following diagram commutes:
\[
\xymatrix{K_0(C^*(\Gamma)) \ar[r]^{R^{(\Gamma)}_Q} \ar[d]_{\cong} & K_0(\C) \ar[d]^{\cong} \\
K^0(\hat \Gamma) \ar[r]^{i^*} & K^0(\{1\}).
}
\]
That is,
\[
R^{(\Gamma)}_Q\left([E]\right) = \dim E_1 = \mathrm{rank}(E) \in \Z,
\]
for all vector bundles $E \to \hat \Gamma$.
\end{lemma}
The proof is a straightforward verification.
\medskip

\textit{End of proof of Proposition \ref{prop Valette ab}.}
From Lemma \ref{lem reduction} and Proposition \ref{prop im mu}, we conclude that
\[
R^{(\Gamma)}_Q \circ \mu^{\Gamma}_M\left[L^2(M, E), F\right] = [\ker D^+_1] - [\ker D^-_1] = \ind D_1 \in \Z.
\]
The Hilbert space $H_1$ is isomorphic to $L^2(M/\Gamma, E/\Gamma)$, and this isomorphism intertwines $D_1$ and $D^{\Gamma}$. 
Hence  Proposition \ref{prop Valette ab} follows.
\hfill $\blacksquare$

\section{Example: action of $\Z^{2n}$ on $\R^{2n}$} \label{sec example}

Let $M$ be the manifold
$M = T^*\R^n \cong \R^{2n} \cong \C^n$.
An element of $M$ is denoted by
$(q, p) := (q_1, p_1, \ldots, q_n, p_n)$,
where $q_j, p_j \in \R$, or by
$q+ip = z := (z_1, \ldots, z_n)$,
where $z_j = q_j + ip_j \in \C$. We equip $M$ with the standard symplectic form
$\omega := \sum_{j=1}^n dp_j \wedge dq_j. $

Let $\Gamma$ be the group
$\Gamma = \Z^{2n} \cong \Z + i\Z.$
The action of $\Gamma$ on $M$ by addition is denoted by $\alpha$.  
Our aim is to find a prequantisation for this action and the corresponding Dirac operator for general $n$, and the quantisation of this action for $n=1$. 
This is less trivial than it may seem because of the coupling of the standard Dirac operator to the prequantum line bundle, which precludes the use of the 
standard formulae. We will then see,
just as in Section \ref{Abeliancase}, that the reduction of this quantisation is the quantisation of the reduced space $\mathbb{T}^2$.

\subsection{Prequantisation}

Let $L := M \times \C \to M$ be the trivial line bundle. Inspired by the construction of line bundles on tori with a given Chern class 
(see e.g.\ \cite{griffithsharris}, pp. 307--317), 
we lift the action of $\Gamma$ on $M$ to an action of $\Gamma$ on $L$ (still called $\alpha$), by setting
\[
\begin{split}
e_j \cdot (z, w) &= (z+e_j, w); \\
ie_j \cdot (z, w) &= (z+ie_j,e^{-2\pi i z_j} w). 
\end{split}
\]
Here $z \in M$, $w \in \C$, and
\[
e_j := (0,\ldots, 0,1,0, \ldots,0) \in \Z^n,
\]
the $1$ being in the $j^{\mathrm{th}}$ place. The corresponding representation of $\Gamma$ in the space of smooth sections of $L$ is denoted by $\rho$:
\[
\left(\rho_{k+il}s\right)(z) = \alpha_{k+il}s(z-k-il),
\]
for $k, l \in \Z^n$ and $z \in M$.
Define the metric $H$ on $L$ by
\[
H\left((z, w), (z, w')\right) = h(z)w\bar w',
\]
where $z \in M$, $w, w' \in \C$, and $h \in C^{\infty}(M)$ is defined by
\[
h(q+ip) := e^{2\pi \sum_{j}(p_j - p_j^2)}. 
\]
Let $\nabla$ be the connection on $L$ defined by
\[
\nabla := d + 2\pi i \sum_{j=1}^n p_j \, dz_j + \pi \, dp_j. 
\]
\begin{proposition}
The triple $(L, H, \nabla)$ defines an equivariant prequantisation for $(M, \omega)$. 
\end{proposition}
The proof of this proposition is a set of tedious computations. Because of the term $2\pi i \sum_{j=1}^n p_j \, dq_j$ in the expression for the connection $\nabla$, it
has the right curvature form. The terms $-2\pi\sum_{j=1}^n   \, p_j \, dp_j$ and $\pi \, dp_j$ do not change the curvature, and have been added to make $\nabla$ 
equivariant. At the same time, the
latter two terms ensure that there is a $\Gamma$-invariant metric (namely $H$) with respect to which $\nabla$ is Hermitian.

As we mentioned in Subsection \ref{sec group actions}, there is a procedure in \cite{hawkins} to lift the action of $\Z^{2n}$ on $\R^{2n}$ to a projective action on $L$ that leaves the connection (for example) $\nabla' := d + 2\pi i \sum_j p_j \, dq_j$ invariant. This projective action turns out to be an actual action in this case, and preserves the standard metric on $L$. We thus obtain prequantisation of this action that looks much simpler than the one given in this section. However, we found our formulas to be 
more suitable to compute the kernel of the associated Dirac operator.

\subsection{The Dirac operator}

In this subsection, we compute the Dolbeault--Dirac operator $\D$ on $M$, coupled to $L$. To compute the quantisation of the action we are considering, we need to compute the
kernels of
\[
\begin{split}
\DS^+ &:= \D|_{\Omega^{0, \mathrm{even}}(M)}; \\
\DS^- &:= \D|_{\Omega^{0, \mathrm{odd}}(M)}.
\end{split}
\]
This is not easy to do in general. But for $n=1$, these kernels are computed in Subsection \ref{sec n=1}.

In our expression for the Dirac operator, we will use multi-indices
\[
l = (l_1, \ldots, l_q) \subset \{1, \ldots, n\},
\]
where $q \in \{ 0, \ldots, n\}$ and $l_1 < \cdots < l_q$. We will write
$d\bar z^l := d\bar z_{l_1} \wedge \ldots \wedge d\bar z_{l_q}$. If $l = \emptyset$, we set $d\bar z^l := 1_M$, the constant function $1$ on $M$.
Note that $\{d\bar z^l\}_{l \subset \{1, \ldots ,n\}}$ is a $C^{\infty}(M)$-basis of $\Omega^{0,*}(M; L)$.

Given $l \subset \{1, \ldots, n\}$ and $j \in \{1, \ldots, n\}$, we define
\[
\varepsilon_{jl}:= (-1)^{\#\{ r \in \{1, \ldots , q\}; l_r < j\}},
\]
plus one if an even number of $l_r$ is smaller than $j$, and minus one if the number of such $l_r$ is odd.
From the definition of the Dolbeault--Dirac operator one then deduces:
\begin{proposition} \label{prop Dirac op on C^n}
For all $l\subset \{1, \ldots , n\}$ and all $f \in C^{\infty}(M)$, we have
\begin{equation} \label{eq dirac op C^n}
\begin{split}
\D\bigl(fd\bar z^l\bigr) &= \sum_{j \in l} \varepsilon_{jl}\left(-2 \frac{\partial f}{\partial z_j} + (i\pi - 4\pi i p_j)f\right) d\bar z^{l \setminus \{j\}}  \\
	& \qquad + \sum_{ \substack{1\leq j \leq n, \\ j \not\in l}} \varepsilon_{jl}\left(\frac{\partial f}{\partial \bar z_j} + \frac{i\pi}{2}f\right) 
		d\bar z^{l \cup \{j\}}. 
\end{split}
\end{equation}
\end{proposition}

\subsection{The case $n=1$} \label{sec n=1}

We now consider the case where $n=1$. That is, $M = \C$ and $\Gamma = \Z + i\Z$. We can then explicitly compute the quantisation
of the action. This will allow us to illustrate our noncompact Guillemin--Sternberg conjecture by computing the four corners in diagram \eqref{qcr}.

If $n=1$, Proposition \ref{prop Dirac op on C^n} reduces to
\begin{corollary}
The Dirac operator on $\C$, coupled to $L$, is given by
\[
\D(f_1 + f_2 d\bar z) = \left(\frac{\partial f_1}{\partial \bar z} + \frac{i\pi}{2} f_1\right) d\bar z -2\frac{\partial f_2}{\partial z} +(i\pi -4\pi i \, p)f_2.
\]
That is to say, with respect to the $C^{\infty}(M)$-basis $\{1_M, d\bar z\}$ of $\Omega^{0,*}(M; L)$, the Dirac operator $\D$ has the matrix form \eqref{Dmatrix}, 
where
\[
\begin{split}
\DS^+ &= \frac{\partial}{\partial \bar z} + \frac{i\pi}{2}; \\
\DS^- &= -2\frac{\partial}{\partial z} + i\pi -4\pi i \, p.
\end{split}
\]
\end{corollary}

In this case, the kernels of $\DS^+$ and $\DS^-$ can be determined explicitly:
\begin{proposition} \label{prop kernels}
The kernel of $\DS^+$ consists of the sections $s$ of $L$ given by
\[
s(z) = e^{-i\pi \bar z/{2}} \varphi(z),
\]
where $\varphi$ is a holomorphic function.

The kernel of $\DS^-$ is isomorphic to the space of smooth sections $t$ of $L$ given by
\[
t(z) = e^{i\pi z/2 + \pi |z|^2 -\pi z^2/2} \overline{\psi(z)},
\]
where $\psi$ is a holomorphic function.
\end{proposition}

The unitary dual of the group $\Z+i\Z = \Z^2$ is the torus $\mathbb{T}^2$. Therefore, by Proposition \ref{prop im mu}, 
the quantisation of the action of $\Z + i\Z$ on $\C$ is the class in \mbox{$K\! K(\C, C^*(\Z^2))$}  that
corresponds to the class
\[
\left[\left(\ker \DS^+_{(\alpha, \beta)}\right)_{(\alpha, \beta) \in \mathbb{T}^2}\right] - 
	\left[\left(\ker \DS^-_{(\alpha, \beta)}\right)_{(\alpha, \beta) \in \mathbb{T}^2}\right]
\]
in $K^0(\mathbb{T}^2)$.  It will turn out that the kernels of $\DS^+_{(\alpha, \beta)}$ and $\DS^-_{(\alpha, \beta)}$ indeed define vector bundles over
$\mathbb{T}^2$.
Let us compute these kernels.
\begin{proposition} \label{prop ker D+ alpha beta}
Let $\lambda, \mu \in \R$. Define the section $s_{\lambda\mu} \in C^{\infty}(M, L)$ by
\[
s_{\lambda \mu}(z) = e^{i\lambda z} e^{-\pi p} \sum_{k \in \Z} e^{-\pi k^2} e^{-k(\lambda + i\mu + 2\pi)} e^{2\pi i k z}.
\]
Set $\alpha  := e^{i\lambda}$ and $\beta := e^{i\mu}$.
Then
$\ker \DS^+_{(\alpha, \beta)} = \C s_{\lambda \mu}$.
\end{proposition}
\begin{remark} \label{rem s alpha beta}
For all $\lambda, \mu \in \R$, we have
\[
\begin{split}
s_{\lambda + 2\pi, \mu} &= e^{\lambda + i\mu + 3\pi} s_{\lambda \mu}; \\
s_{\lambda, \mu + 2\pi} &= s_{\lambda \mu}.
\end{split}
\]
Hence the vector space
$\C s_{\lambda \mu} \subset C^{\infty}(M, L)$
is invariant under $\lambda \mapsto \lambda + 2\pi$ and $\mu \mapsto \mu + 2\pi$. 
This is in agreement with the fact that $\C s_{\lambda \mu}$ is the kernel of $\DS^+_{(e^{i\lambda}, e^{i\mu})}$. 
\end{remark}
\textit{Sketch of proof of Proposition \ref{prop ker D+ alpha beta}.}
Let $\lambda, \mu \in \R$, and $s \in C^{\infty}(M, L) = C^{\infty}(\C, \C)$. Suppose $s$ is in the kernel of $\DS^+_{(\alpha, \beta)}$. Let $\varphi$ be the holomorphic
function from Proposition \ref{prop kernels}, and write
\[
\tilde \varphi(z) := e^{-i\lambda z}e^{-i\pi z/2} \varphi(z) = \sum_{k \in \Z} a_k \, e^{2\pi i k z}
\]
(note that for all $z \in \C$ one has $\tilde \varphi(z+1)= \tilde \varphi(z)$).
Then
$a_k = e^{-\pi k^2} e^{-k(\lambda + i \mu + 2\pi)} a_0$,
which gives the desired result.
\hfill $\blacksquare$

\begin{proposition}
The kernel of $\DS^-_{(\alpha, \beta)}$ is trivial for all $(\alpha, \beta) \in \mathbb{T}^2$.
\end{proposition}
\noindent\textit{Sketch of proof.}
Let $\lambda, \mu \in \R$ and let $t\, d\bar z \in \Omega^{0,1}(M; L) = C^{\infty}(M, L) d\bar z$. Suppose that $t \, d\bar z \in \ker \DS^-_{(e^{i\lambda}, e^{i\mu})}$. 
Let $\psi$ be the holomorphic function from  Proposition \ref{prop kernels}, and write
\[
\tilde \psi(z) := e^{\pi(\bar z^2 + i\bar z)/2 - i\lambda \bar z} \overline{\psi(z)} = \sum_{k \in \Z} c_k \, e^{2\pi i k \bar z}
\]
(note that for all $z \in \C$ one has $\tilde \psi(z+1)= \tilde \psi(z)$).
Then
$c_k = e^{\pi k^2} e^{k(\lambda -i\mu -2\pi)} c_0$,
which implies that $c_0=0$.
\hfill $\blacksquare$
\medskip

We conclude:
\begin{proposition}
The quantisation of the action of $\Z^2$ on $\C$ is the class in $K^0(\mathbb{T}^2)$ defined by the vector bundle\footnote{By 
Remark \ref{rem s alpha beta}, this is indeed a well-defined vector bundle.}
\[
\left(\C s_{\lambda \mu} \right)_{(e^{i\lambda}, e^{i \mu}) \in \mathbb{T}^2} \to \mathbb{T}^2.
\]
\end{proposition}

By Lemma \ref{lem reduction}, we now find that  the reduction of the quantisation of the action of $\Z^2$ on $\R^2$ is the
one-dimensional vector space
$\C \cdot s_{0,0} \subset C^{\infty}(M, L)$,
where
\[
s_{0,0}(z) = e^{-\pi p} \sum_{k \in \Z} e^{-\pi k^2} e^{-2\pi k} e^{2\pi i k z}.
\]
As we saw in Section \ref{sec ass map abelian}, this is precisely the index of $\DS_{\mathbb{T}^2}^{L/\Z^2}$.
Schematically, we therefore have\footnote{
Note that it is a coincidence that the two-torus appears twice in this diagram: in this example $M/\Gamma = \mathbb{T}^2 = \hat \Gamma$.}
\[
\xymatrix{
\Z^2 \circlearrowright \R^2 \ar@{|->}[r]^-Q  \ar@{|->}[d]_{R_C} & \left(\C s_{\lambda \mu} \right)_{(e^{i\lambda}, e^{i \mu}) \in \mathbb{T}^2} \ar@{|->}[d]^{R_Q} \\
\mathbb{T}^2 \ar@{|->}[r]^-Q & \C \cdot s_{0,0}.
}
\]

\begin{remark}
The fact that the geometric quantisation of the torus $\mathbb{T}^2$ is one-dimensional can alternatively be deduced from
 the Atiyah-Singer index theorem for Dirac operators. 
Indeed, let $\DS_{\mathbb{T}^2}^{L/\Z^2}$ be the Dirac operator on the torus, coupled to the quotient line bundle $L/\Z^2$. Then by Atiyah-Singer, in the form  stated 
for example in \cite{GGK} on page 117, one has
\[
\begin{split}
Q(\mathbb{T}^2) = \ind \DS_{\mathbb{T}^2}^{L/\Z^2} &= \int_{\mathbb{T}^2} e^{\mathrm{ch}_1(L/\Z^2)} \\
	&= \int_{\mathbb{T}^2} dp \wedge dq \\
	&=1,
\end{split}
\]
the symplectic volume of the torus, i.e.\ the volume determined by the Liouville measure.
\end{remark}

\appendix

\section{Naturality of the assembly map} \label{app nat}
Our ``quantisation commutes with reduction'' result is partly a consequence of the naturality of the assembly map. For discrete groups, this naturality is explained in detail by Valette
\cite{valette}. We need to generalise ``one half'' of this naturality (the epimorphism case) to non-discrete groups. 

\subsection{The statement}
Let $G$ be a locally compact unimodular group acting properly on a locally compact Hausdorff space $X$. We consider a closed normal 
subgroup\footnote{Although for our purpose it is enough to consider \textit{discrete} normal subgroups of $G$, we have to work in the nondiscrete setting anyway, since $G$ is not necessarily discrete. We therefore allow nondiscrete subgroups $N$.} 
 $N$ of $G$, and suppose that $G$ and $N$ are equipped with Haar measures $dg$ and $dn$, respectively. 
We suppose that $X/G$ is compact.

The version of naturality of the assembly map that we will need is the following.
\begin{theorem} \label{thm nat}
The homomorphism $V_N$ defined in Subsection \ref{subsec L^2} makes the following diagram commute:
\[
\xymatrix{
K_0^G(X) \ar[r]^{\mu^G_X} \ar[d]_{V_N} & K_0(C^*(G)) \ar[d]^{R_Q^{(N)}} \\
K_0^{G/N}(X/N) \ar[r]^{\mu^{G/N}_{X/N}} & K_0(C^*(G/N)).
}
\] 
\end{theorem}
Here $\mu^G_X$ and $\mu^{G/N}_{X/N}$ are analytic assembly maps as defined in e.g.\ \cite{BCH, Val1, valette}, and the map 
\beq R_Q^{(N)}= \left({\textstyle \int_N}\right)_*\eeq
 is functorially induced by the map $\int_N: C^*(G) \to C^*(G/N)$ given on $f\in C_c(G)$ 
by \cite{Green77}
\beq {\textstyle \int_N} (f): Ng \mapsto  \int_N f(ng)\, dn.\eeq 

To prove Theorem \ref{thm nat}, one can simply copy the proof for discrete groups in Valette \cite{valette}, 
replacing discrete groups by possibly non-discrete ones and sums by integrals. In places where Valette uses the fact that a finite sum of compact operators is again compact, one uses Lemma \ref{lem int cpt a}. This lemma states that in some cases, the integral over a compact set of a continuous family of compact operators is compact. 

Another difference between the discrete and the nondiscrete cases is the definition of the map $\Psi$ on page 110 of \cite{valette}. In the nondiscrete case this map is defined as follows. If $\xi \in H_c$, we will write $\xi^N := \xi + \ker (\cdot, \cdot)_N$ for its class in $H_N$. Then for all $\xi \in H_c$, we have $\xi^N \in H_{N, c}$. 
Define the linear map
\[
\Psi: H_c \otimes_{C_c(G)} C_c(G/N) \to H_N
\]
by
\[
\Psi[\xi \otimes \varphi] = \int_{G/N} \varphi(Ng^{-1}) \, Ng \cdot \xi^N  d(Ng),
\]
where $d(Ng)$ is the Haar measure on $G/N$ corresponding to the Haar measures $dn$ and $dg$ on $N$ and $G$, respectively. 
To prove that the extension $\Psi: \mathcal{E} \otimes_{C^*(G)}C^*(G/N) \to \tilde{\mathcal{E}}$ is surjective, one can use a sequence of compactly supported continuous functions on $G/N$ that converges to the distribution $\delta_{Ne} \in \mathcal{D}'(G/N)$ with respect to the measure $d(Ng)$.

\subsection{Integrals of families of operators} \label{sec int ops}

\begin{lemma} \label{lem int cpt a}
Let $\E$ be a Hilbert $C^*$-module, and let $\F(\E)$ and $\K(\E) = \overline{\F(\E)}$ be
the algebras of finite-rank and compact operators on $\E$, respectively.
Let $(M,\mu)$ be a 
compact Borel space with finite measure. Suppose $M$ is metrisable. Let  
$\alpha, \beta: M \to \B(\E)$
be continous, and let $T \in \K(\E)$ be a compact operator. Define the map
$\phi: M \to \K(\E)$
by
$\phi(m) = \alpha(m)T\beta(m)$.
Then the integral $\int_M \phi(m)\, dm$ defines a compact operator on $\E$.
\end{lemma}
We will prove this lemma in several steps. For continuous maps $\psi: M \to \mathcal{B}(\E)$, we will use the norm
\[
\|\psi\|_{\infty} := \sup_{m \in M} \|\psi(m) \|_{\B(\E)}.
\]

\begin{lemma} \label{lem int cpt}
Let $(M,\mu)$ be a compact Borel space with finite measure, and let $\E$ be a Hilbert $C^*$-module.
Let 
$\phi: M \to \K(\E)$
be a continuous map. Suppose that $\phi$ is `uniformly compact', in the sense that there exists a sequence
$\bigl(\phi_j\bigr)_{j=1}^{\infty}: M \to \F(\E)$
such that
$\|\phi_j - \phi\|_{\infty}$
tends to zero as $j \to \infty$. Suppose furthermore that for every $j \in \N$, there is a sequence
$\bigl(\phi_j^k\bigr)_{k=1}^{\infty}:M \to \F(\E)$
of simple functions (i.e.\ having finitely many values), such that for all $\varepsilon>0$ there is an $n \in \N$ such that for all $j,k \geq n$,
$\|\phi_j^k - \phi_j\| < \varepsilon$.
Then the integral
$\int_M \phi(m) d\mu(m)$
defines a compact operator on $\E$.
\end{lemma}
\begin{proof}
For all $j, k \in \N$, the integral
$\int_M \phi_j^k(m) d\mu(m)$
is a finite sum of finite-rank operators, and hence a finite-rank operator itself. And because
$\|\phi_j^j - \phi \|_{\infty} \to 0$
as $j$ tends to $\infty$, we have
\[
\int_M \phi_j^j(m) d\mu(m) \to \int_M \phi(m) d\mu(m)
\]
in $\B(\E)$. Hence $\int_M \phi(m) d\mu(m)$ is a compact operator.
\end{proof}

\begin{lemma} \label{lem ass cpt}
In the situation of Lemma \ref{lem int cpt a}, the conditions of
 Lemma \ref{lem int cpt} are satisfied..
\end{lemma}
\begin{proof}
 Choose a sequence $(T_j)_{j=1}^{\infty}$ in $\F(\E)$ that
converges to $T$. For $m \in M$, set
\[
\phi_j(m) = \alpha(m)T_j\beta(m)
\]
Then
\[
\|\phi_j - \phi\|_{\infty} \leq \|\alpha\|_{\infty} \|T_j - T\|_{\B(\E)} \|\beta\|_{\infty} \to 0
\]
as $j \to \infty$. Note that $\alpha$ and $\beta$ are continuous functions on a compact space, so their $\sup$-norms are bounded.

Choose sequences of simple functions
$\alpha^k, \beta^k: M \to \B(\E)$
such that $\|\alpha^k - \alpha\|_{\infty} \to 0$ and $\|\beta^k - \beta\|_{\infty} \to 0$ as $j$ goes to $\infty$ (see Lemma \ref{lem simple} below).
For all $j,k \in \N$, set
\[
\phi_j^k(m) := \alpha^k(m)T_j\beta^k(m),
\]
for $m \in M$. Note that
\[
\begin{split}
\|\phi_j^k - \phi_j\|_{\infty} &= \sup_{m \in M} \|\alpha^k(m) T_j \beta^k(m) - \alpha(m)T_j\beta(m)\| \\
	&= \sup_{m \in M} \Bigl( \|\alpha^k(m) T_j \beta^k(m) - \alpha^k(m)T_j\beta(m)\|   \\
	& \qquad  \qquad +  \|\alpha^k(m) T_j \beta(m) - \alpha(m)T_j\beta(m)\| \Bigr) \\
	&\leq \|\alpha^k\|_{\infty} \|T_j\| \|\beta^k - \beta\|_{\infty} + 
		\|\alpha^k - \alpha\|_{\infty} \|T_j\| \|\beta\|_{\infty}.
\end{split}
\]
The sequences $k \mapsto \|\alpha^k\|_{\infty}$ and $j \mapsto \|T_j\|$ are bounded, since $\alpha^k \to \alpha$ and $T_j \to T$. Hence, because the sequences $\|\alpha^k -
\alpha\|_{\infty}$ and $\|\beta^k - \beta\|_{\infty}$ tend to zero, we see that $\|\phi_j^k - \phi_j\|$ can be made smaller than any $\varepsilon > 0$ 
for $k$ large enough, uniformly in $j$.
\end{proof}

\begin{lemma} \label{lem simple}
Let $(M, \mu)$ be a metrisable compact Borel space with metric $d_M$, let $Y$ be a normed vector space, and let
$\alpha: M \to Y$
be a continuous map.

Then there exists a sequence of simple maps
$\alpha^k: M \to Y$
such that $\bigl(\alpha^k\bigr)^{-1}(V)$ is measurable for all $V \subset Y$, and such that the sequence
$\bigl(\|\alpha - \alpha^k\|_{\infty}\bigr)_{k=1}^{\infty}$
goes to zero as $k$ goes to infinity.
\end{lemma}
\begin{proof}
For every $k \in \N$, choose a finite covering
$\tilde U_k = \{\tilde V_k^1, \ldots, \tilde V_k^{n_k}\}$
of $M$ by balls of radius $\frac{1}{k}$. From each $\tilde U_k$, we construct a partition 
$U_k = \{V_k^1, \ldots, V_k^{n_k}\} $
of $M$, by setting $V_k^1 := \tilde V_k^1$, and
$V_k^j := \tilde V_k^j \setminus \bigcup_{i=1}^{j-1}\tilde V_k^i$.
for $j=2, \ldots, n_k$. Note that the sets $V_k^j$ are Borel-measurable.
For all $k \in \N$ and $j \in \{1, \ldots, n_k\}$, choose an element $m_k^j \in V_k^j$. Define the simple map
$\alpha^{k}: M \to Y$
by
\[
\alpha^k(m) := \alpha(m_k^j) \qquad \text{if $m \in V_k^j$.}
\]

Note that, because $\alpha$ is continuous (and uniformly continuous because $M$ is compact), for every $\varepsilon > 0$ there is 
a $k_{\varepsilon} \in \N$ such that for all $m, n \in M$,
\[
d_M(m, n) < \frac{1}{k_{\varepsilon}} \quad \Rightarrow \quad \|\alpha(m) - \alpha(n)\|_Y < \varepsilon.
\]
Hence for all $\varepsilon > 0$, all $k > k_{\varepsilon}$, and all $m \in M$ (say $m \in V_k^j$),
\[
\|\alpha(m) - \alpha^k(m)\|_Y = \|\alpha(m) - \alpha(m_k^j)\|_Y < \varepsilon.
\]
So $\|\alpha - \alpha^k\|_{\infty}$ indeed goes to zero.
\end{proof}

\end{document}